\documentclass[journal]{IEEEtran}

\IEEEoverridecommandlockouts                              

\usepackage{graphics} 
\usepackage{epsfig} 
\usepackage{amsmath} 
\usepackage[hidelinks]{hyperref}

\usepackage{svg}

\usepackage{amssymb}  
\usepackage{pifont}
\usepackage{multirow}
\usepackage{booktabs}
\usepackage{cite}
\usepackage{tabularx}
\usepackage{comment}
\usepackage{subcaption}
\usepackage{lipsum} 
\usepackage{graphicx,subcaption}
\usepackage[font={footnotesize,it}]{caption}
\usepackage{multirow}
\usepackage{algorithm}

\usepackage[noend]{algpseudocode}

\definecolor{blue}{RGB}{0,0,0}
\definecolor{blue2}{RGB}{0,0,0}
\definecolor{blue3}{RGB}{0,0,0}

\usepackage{fancyhdr}
\usepackage{ragged2e}
\fancypagestyle{ieee_notice}
{
    \setlength{\headheight}{32pt}
    \addtolength{\topmargin}{-5pt}
    \lhead{\fbox{\begin{minipage}{\textwidth}
    \noindent\footnotesize\justifying\copyright 2024 IEEE. Personal use of this material is permitted. Permission from IEEE must be obtained for all other uses, in any current or future media, including reprinting/republishing this material for advertising or promotional purposes, creating new collective works, for resale or redistribution to servers or lists, or reuse of any copyrighted component of this work in other works. This is a preprint of the paper accepted to T-Ro.\end{minipage}}}
    \cfoot{} 
}

\DeclareMathOperator{\sgn}{sgn}

\makeatletter
\newcommand{\thickhline}{%
    \noalign {\ifnum 0=`}\fi \hrule height 1pt
    \futurelet \reserved@a \@xhline
}
\newcolumntype{"}{@{\hskip\tabcolsep\vrule width 1pt\hskip\tabcolsep}}
\makeatother

\title{\LARGE \bf Planar Friction Modelling with LuGre Dynamics and Limit Surfaces}

\author{Gabriel Arslan Waltersson and Yiannis Karayiannidis

\thanks{This work was funded by the Wallenberg AI, Autonomous Systems and Software Program (WASP) funded by the Knut and Alice Wallenberg Foundation.}
\thanks{Gabriel Arslan Waltersson is  with the Department of Electrical Engineering, Chalmers University of Technology, SE-412 96 Gothenburg, Sweden 
         {\tt\small gabwal@chalmers.se}. 
 {\textit{Corresponding author.}}}
        \thanks{Yiannis Karayiannidis is with the Department of Automatic Control, Lund University, Sweden {\tt\small  yiannis@control.lth.se}. The author is a member of the ELLIIT Strategic Research Area at Lund University. }

}

\begin{document}

\maketitle
\thispagestyle{ieee_notice} 
\pagestyle{empty}


\begin{abstract}
\textcolor{blue3}{During planar motion, contact surfaces} exhibit a coupling between tangential and rotational friction forces. This paper proposes planar friction models grounded in the LuGre model and limit surface theory. First, distributed planar extended state models are proposed and the Elasto-Plastic model is extended for multi-dimensional friction. Subsequently, we derive a reduced planar friction model coupled with a pre-calculated limit surface, that offers reduced computational cost. The limit surface approximation through an ellipsoid is discussed. The properties of the planar friction models are assessed in various simulations, demonstrating that the reduced planar friction model achieves comparable performance to the distributed model while exhibiting $\sim 80$ times lower computational cost. 
\end{abstract}

\section{Introduction}

\begin{figure}
    \centering
    \smallskip 
    \includegraphics[width=0.9\columnwidth]{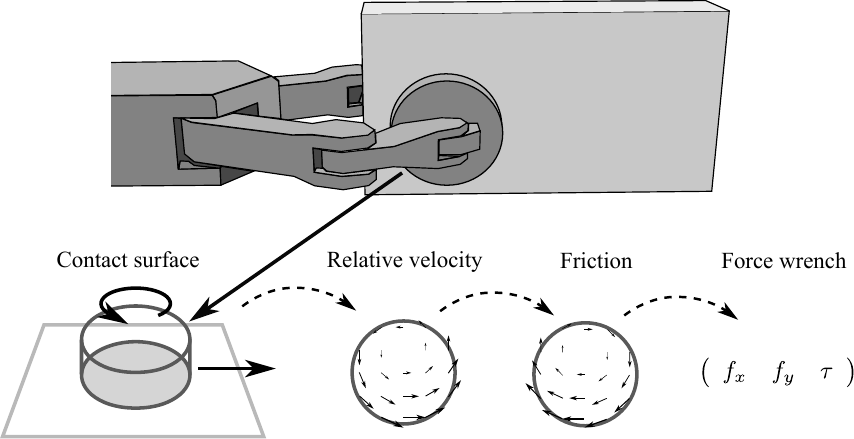}
    \caption{Planar friction in a parallel gripper, wherein the friction force is represented by a force wrench vector.}
    \label{fig:Overview}
    \vspace*{-0.5cm}
\end{figure}

Humans possess the remarkable ability to engage in intricate manipulation and interaction with the world around us. We depend on various sensory channels, including vision, touch, and hearing, among others, to perceive our surroundings. Within the realm of robotics, vision has traditionally served as the predominant mode of perception \cite{Natale_2017}. However, for navigation in human environments or in-hand object manipulation, there exists valuable information that vision alone cannot easily discern, such as object weight, friction properties, applied forces, or grasp state. In such cases, tactile and haptic perception \textcolor{blue3}{are proven to be} more suitable \cite{KEMP2007Challenges}. The interaction between a robot and its environment hinges upon the principles of friction and contact dynamics, e.g. an object gripped by a parallel gripper, or a walking robot's foot making contact with the ground. \textcolor{blue2}{There has been a noticeable gap in planar friction models that are capable of capturing the critical behaviour in the slip-stick regime, similar to what one DoF models can achieve.}
Therefore, this paper focuses on modelling \textcolor{blue2}{dynamic} friction for arbitrary contact surfaces in planar motion, \textcolor{blue3}{with} an example depicted in Fig. \ref{fig:Overview}. This paper addresses several challenges encountered in dynamic planar friction models, including computational cost, numerical approximation, and drifting under oscillating loads.

The study of friction encompasses various disciplines \cite{berger2002friction}, each of which tends to possess its own set of friction models tailored to its specific requirements. Planar friction exists in many fields. In robotics, it is found in the context of in-hand manipulation, walking/ground contact, pushing, etc. In the domain of vehicles, planar friction becomes relevant in areas such as tire modelling and safety systems e.g. traction control or ABS. In recent times, with the emergence of digital twins and more sophisticated physics simulators like PhysX used by Isaac Gym \cite{makoviychuk2021isaac} or Algoryx \cite{lacoursiere2015framework}, the relevance of these models has increased. \footnote{The code developed for this paper is available with open access at \url{https://github.com/Gabrieleenx/FrictionModelling}.}  

\section{Related work}
\begin{figure}
    \centering
    \smallskip 
    \includegraphics[width=\columnwidth]{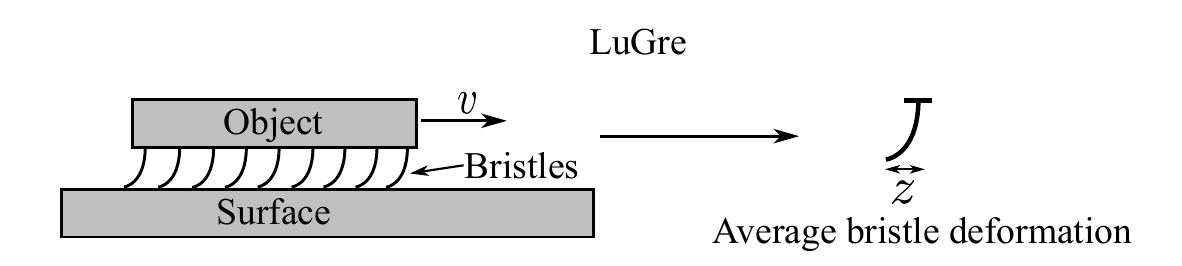}
    \caption{The bristle analogy of the Dahl and LuGre friction model.}
    \label{fig:LuGreBristles}
    \vspace*{-0.5cm}
\end{figure}

The classical Coulomb friction model describes friction as a force  that opposes the motion and is proportional to the normal force, or a force that in static scenarios balance the applied forces acting on the body.
Dahl introduced a friction model in 1968 \cite{Dahl1968ASF}, which estimates friction forces based solely on velocity and micro displacements by incorporating an additional state variable. The model conceptualizes the friction force as a result of numerous tiny bristles present at the contact surface. The extended state variable \textcolor{blue3}{represents} the average deflection of the bristles, \textcolor{blue3}{as} illustrated in Fig. \ref{fig:LuGreBristles}. However, Dahl's model does not account for the Stribeck effect \cite{LuGre2008}, encoding the dependency of the friction coefficient and viscosity on velocity \cite{Stribeck1902}. To address this limitation, Åström and Canudas-de-wit developed the LuGre model, which extends the Dahl friction model to incorporate the Stribeck effect \cite{LuGre2008}. 

Over time, the LuGre model has undergone numerous extensions and modifications to enhance its capabilities. One of the early modifications was the inclusion of varying normal force, resulting in the amended LuGre model \cite{de1999dynamic}. The original LuGre model exhibited drifting behaviour for varying tangential loads, even when the load remained below the maximum static friction threshold. To tackle this issue, Dupont et al. proposed the Elasto-Plastic model \cite{dupont2000elasto} as an extension of the LuGre model. In the Elasto-Plastic model, the internal state variable representing bristle deflection behaves as a pure spring for deflections smaller than the break-away deflection. 
In experimental comparisons between Coulomb, Stribeck, Dahl, LuGre and Elasto-Plastic friction models for micro stick-slip motions, Liu et al. found that the LuGre model exhibited the highest level of accuracy \cite{Lui2015ExperimentalFriction}.
Marques et al. conducted investigations into the LuGre model under varying normal loads \cite{marques2021investigation}. They observed that during stiction, the amended LuGre model experienced drifting in displacement and oscillations in tangential force. These issues were attributed to the varying bristle deflection required to maintain a constant tangential force. To mitigate the oscillations, the proposed solution \textcolor{blue3}{introduced}
a variable bristle stiffness coefficient that remained fixed during sticking. However, this approach resulted in a trade-off, as the model's stiffness was no longer consistent and depended on how the sticking state was established. In Section \ref{sec:results_drifing}, we present results demonstrating that the Elasto-Plastic model can alleviate drifting in displacements for varying normal loads. The LuGre model can introduce challenges in simulating dynamical systems due to its numerical stiffness\textcolor{blue3}{. Recognizing} this issue, Do \emph{et al.} \cite{do2007efficient} studied different methods for simulating systems with the LuGre model. To enhance computational efficiency in simulating multi-zone contacts of thin beams, Wang utilized a piecewise analytical LuGre model \cite{wang2016dynamic}. \textcolor{blue2}{Rill \emph{et al.} \cite{RillLuGre2023} recently introduced a friction model with second-order bristle dynamics and compared the LuGre implementation \textcolor{blue3}{with Adams and Matlab simulators.}}

The friction models discussed thus far have been one-dimensional. However, when dealing with planar motion, friction becomes multi-dimensional. In the context of planar sliders, the limit surface theory, introduced by Goyal et al. \cite{goyal1989limit} \cite{goyal1991planar}, is \textcolor{blue3}{a} prominent theory that describes the coupling between tangential friction and torque for different motions, assuming Coulomb friction. The limit surface theory has gained interest in the robotics community and in modelling contact mechanics for grasping tasks. The limit surface is often approximated using an ellipsoid, due to its similar shape and simplified representation \cite{howe1996practical, ghazaei2020quasi, xydas1999modeling, fakhari2014dynamic, bicchi1993experimental, shi2017dynamic}. It is worth noting that the comparison between the limit surface and the ellipsoid approximation is commonly depicted using plots that focus solely on the closest distance. However, in Section \ref{sec:ellipse_approximation}, we present evidence to demonstrate that the ellipsoid approximation of the limit surface is less accurate when considering the corresponding location on the limit surface than \textcolor{blue3}{how it} is frequently portrayed in the literature.

Hertz pioneered the study of contact mechanics \cite{hertz1882contact}. He established that the radius of the contact area between a sphere and a flat surface is proportional to the normal force raised to the power of 1/3. Building upon Hertz's work, Xydas and Kao extended the model and conducted experiments that revealed the radius to be proportional to the normal force raised to the power between 0 and 1/3, depending on the material properties \cite{xydas1999modeling}. \textcolor{blue2}{Elandt \emph{et al.} \cite{RyanPFC2019} proposed a pressure field contact model for predicting the contact surface, pressure distribution and contact wrench between rigid objects. Masterjohn \emph{et al.} \cite{Materjohn2022} introduced a discrete in time approximation of the pressure field contact model. In both papers, the tangential friction forces were calculated with Coulomb friction.}

In the context of the onset of sliding, Dahmen \emph{et al.} investigated the coupling between static friction and torque, finding that Coulomb friction alone is insufficient to determine this behaviour \cite{dahmen2005macroscopic}. Their experiments also indicated that the limit surface is similar across different materials.
Shi \emph{et al.} utilized the ellipse approximation in their work on dynamic in-hand sliding manipulation of flat objects \cite{shi2017dynamic}. Alternatively, Hu et al. approached the modelling of general planar contacts by incorporating a random field to represent the friction coefficient \cite{hu2021coefficient}.

Only a few papers have attempted to extend dynamic friction models to planar motions. Velenis, Tsiotras, and Canudas-de-Wit presented an early extension of the LuGre model to 2D for tire friction \cite{velenis2002extension}. Kato \cite{kato2014anisotropic, kato2015friction} worked on anisotropic adhesion models with the Dahl model and extended it for 2D motion, by dividing the contact path into small regions both friction and torque could be calculated. Zhou \emph{et al.} modelled 2D friction using the LuGre model for point contacts \cite{zhou2021modeling}. Costanzo \emph{et al.} combined the LuGre model with the limit surface to estimate instantaneous rotation around the \textcolor{blue3}{center} of rotation (CoR) for in-hand pivoting manipulations \cite{costanzo2019two}. \textcolor{blue2}{In \cite{COSTANZO2021Viscus} Constanzo generalized the method, allowing for friction wrenches outside of the limit surface by considering the viscous friction and improving the accuracy of the tracking. In \cite{LuGre2008} Åström and Canudas de Wit discuss the passivity conditions of the viscus friction in the LuGre model.}  
Colantonio \emph{et al.} projected the LuGre model into 3D, capturing friction in multiple directions, but not torsion directly \cite{colantonio20223d}. An attempt to extend the Elasto-Plastic model for planar motion was done in \cite{rogner2017friction}. In our work, we take a different approach, focusing on the general simulation of planar motions, considering different contact pressures, and leveraging the principles of the limit surface theory. 

The paper proposes a computationally efficient model to simulate dynamic and static friction for contact surfaces in planar motion, e.g. grasping or pushing operations. The concrete contributions are:
\begin{itemize}
\item A distributed planar friction model for an arbitrary contact surface and pressure distribution, combining the LuGre model and the limit surface theory, in contrast to \cite{de1999dynamic}, \cite{kato2015friction} which considered specific contact scenarios and cannot model grasping. 
\item A reduced planar friction model with a pre-computed limit surface, that in comparison to the distributed planar friction model is computationally efficient.
\item Extension of the \textcolor{blue3}{Elasto-Plastic} model, originally defined for 1D systems \cite{dupont2000elasto}, to planar friction.

\end{itemize}


The paper is structured as follows: In \textcolor{blue3}{Section}  \ref{sec:theory}, the background theory for the LuGre, Elasto-Plastic and limit surface \textcolor{blue3}{models} is given. In \textcolor{blue3}{Section} \ref{sec:full_model}, the distributed planar friction model is presented along with the multi-dimensional Elasto-Plastic model and the bristle bound. In \textcolor{blue3}{Section} \ref{sec:reduced_planar_friction}, a computationally fast planar friction model is derived. In \textcolor{blue3}{Section} \ref{sec:results} the tests and results for the friction models are presented.

\section{Background}\label{sec:theory}

\subsection{Notation}

We introduce the following notation:
\begin{itemize}
\item Bold lower case letters denote vectors and bold upper letters denote matrices.
\item $|\cdot|$ \textcolor{blue3}{denotes} absolute value of a scalar and element-wise absolute value of vectors and matrices.
\item $\|\cdot\|$ \textcolor{blue3}{denotes} the Euclidean norm of a vector and $\|\cdot\|_\mathbf{M}$ a weighted Euclidean norm where $\mathbf{M}$ is a positive definite matrix.
\item $\textrm{diag}(\mathbf{a})$ with $\mathbf{a}$ being an $n$-dimensional vector  denotes a square diagonal matrix of dimension $n$ with diagonal entries being equal to to the elements of $\mathbf{a}$.
\item The cross product $\mathbf{a}\times\mathbf{b}$ of two dimensional vectors $\mathbf{a}=[a_x\;a_y]^T$, $\mathbf{b}=[b_x\;b_y]^T$ is given by  $\mathbf{a}\times\mathbf{b}=a_xb_y-a_yb_x$.
\item$\textrm{sgn}(\cdot)$ denotes the sign of a scalar and element-wise sign value of vectors and matrics. 
\end{itemize}

\subsection{LuGre \textcolor{blue3}{model}}

In its original form, the LuGre model assumes \textcolor{blue3}{a} constant normal force ($f_N$). However, De Wit and Tsiotras \textcolor{blue2}{\cite{de1999dynamic}} expanded upon this model to accommodate varying normal loads by introducing \textcolor{blue3}{an} active friction coefficient. The active friction coefficient is multiplied by the normal load to determine the resulting friction forces. This adaptation, known as the amended LuGre model, governs the deflection of the bristles, denoted as $z$, as follows:
\begin{equation}\label{eq:dot_z}
    \dot{z} = v - z \frac{\sigma_0 |v|}{g(v)}
\end{equation}
where $v$ is the relative velocity between the surfaces, and $\sigma_0$ is the amended bristle stiffness. The amended steady-state friction coefficient $g(v)$ is given by:
\begin{equation}\label{eq:G_A}
    g(v) = \mu_C + (\mu_S - \mu_C) e^{-|\frac{v}{v_s}|^{\gamma}}
\end{equation}
where $\mu_C$ represents the Coulomb friction and $\mu_S$ the static friction coefficient. The \textcolor{blue3}{parameters} $v_s$ and $\gamma$ govern how fast $g(v)$ converges to $\mu_C$ with increasing velocity. The friction force is expressed as:
\begin{equation}
    f = (\sigma_0 z + \sigma_1 \dot{z} + f_v(v)) f_N
\end{equation}
where $\sigma_1$ corresponds to the dampening coefficient and $f_v(v)$ \textcolor{blue3}{to} the velocity dependent memoryless friction. Typically, $f_v(v) = \sigma_2 v$ denotes linear viscous friction.  

\begin{figure}
    \centering
    \smallskip 
    \includegraphics[width=0.3\columnwidth]{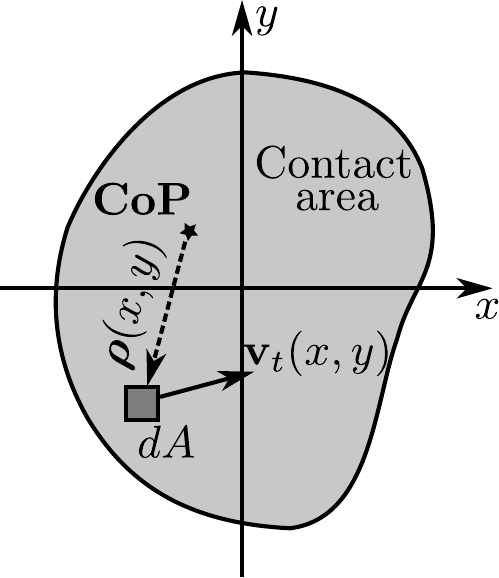}
    \caption{The vector $\boldsymbol{\rho}(x,y)$ from CoP to the infinitesimal contact area $dA$ and the local velocity vector $\mathbf{v}_t(x,y)$, for a contact surface.}
    \label{fig:contactArea}
    \vspace*{-0.5cm}
\end{figure}

\subsection{Elasto-\textcolor{blue3}{P}lastic model}\label{sec:elastPlastic}

It has been observed that the LuGre model exhibits a tendency to drift when subjected to oscillating tangential loads, even when these loads remain significantly below the threshold required for slippage \cite{dupont2000elasto}. To address this issue, the Elasto-Plastic model introduces a purely elastic behaviour for bristle deflections below a specified breakaway threshold, denoted as $z_\textrm{ba}$. The bristle rate deflection \eqref{eq:dot_z} is modified to: 
\begin{equation}\label{eq:elasto_plastic_bristle_deflection}
    \dot{z} = v - \alpha(z, v) z \frac{\sigma_0 |v|}{g(v)}
\end{equation}
where $\alpha(z, v)$ is a function described by:
\begin{equation}\label{eq:elasto_plastic_alpha_fun}
    \alpha(z, v) = \begin{cases} 0 & \sgn(v) \neq \sgn(z) \\  \begin{cases} 0 & |z| \leq z_\textrm{ba} \\ \alpha(z) & z_\textrm{ba} \leq |z| \leq z_\textrm{max}  \\ 1 & |z| \geq z_\textrm{max}  \end{cases} & \sgn(v) = \sgn(z) \end{cases}
\end{equation}
where:
\begin{equation}\label{eq:alpha_z}
    \alpha(z) = \frac{1}{2} \sin \left( \pi \frac{z-(z_\textrm{max} + z_\textrm{ba})/2}{z_\textrm{max }- z_\textrm{ba}}\right) + \frac{1}{2}.
\end{equation}
This sinusoidal function facilitates a smooth transition from elastic to plastic deformation. The elastic behaviour within the sticking regime ensures that the friction force remains proportional to the micro-displacement, thereby preventing drifting. The maximum bristle deflection is determined by the steady-state deflection $z^\star$ and can be expressed as $z_\textrm{max} = z^\star = \frac{ g(v) }{\sigma_0}$.
 
\begin{figure}
    \centering
    \smallskip 
    \includegraphics[width=\columnwidth]{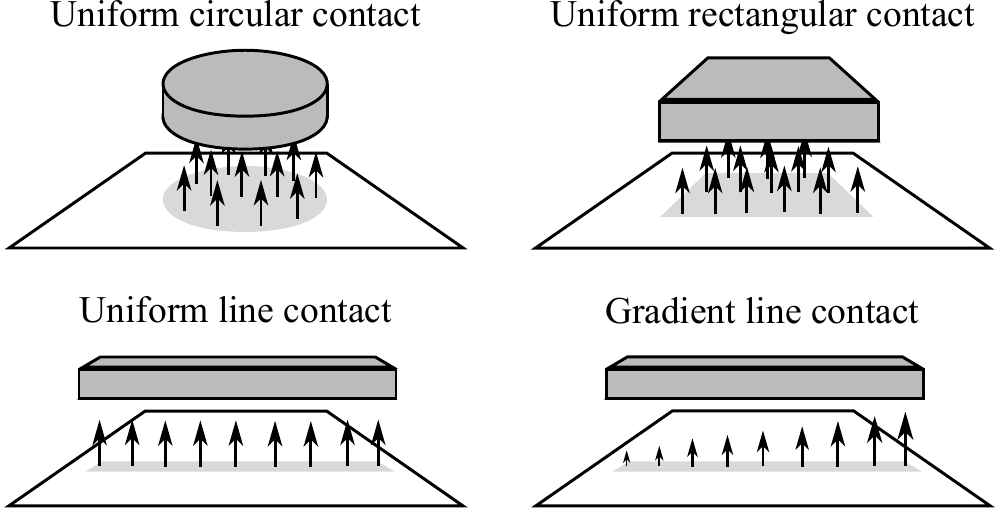}
    \caption{Four illustrations of different contact surfaces.}
    \label{fig:contactPressure}
    \vspace*{-0.5cm}
\end{figure}

\subsection{Limit surface}
A contact surface under planar motion, involving sliding and spinning, has a coupling between tangential and angular friction forces. For example, an object that is spinning requires less force to push it tangentially compared to when the object has no angular velocity. The description of this coupling phenomenon is provided by the limit surface theory \cite{goyal1991planar}, assuming Coulomb friction. The tangential forces $\mathbf{f}^{\textrm{ls}}_{t} \in \Re^2$ and the torque $\tau^{\textrm{ls}}$ defining the limit surface are given by:
\begin{equation}\label{eq:limit_suface_ft}
    \mathbf{f}^{\textrm{ls}}_{t} = -\int_A \mu_C \hat{\mathbf{v}}_t(x,y) p(x,y) d A
\end{equation}

and
\begin{equation}\label{eq:limit_surface_tau}
    \tau^\textrm{ls} = - \int_A \mu_C [\boldsymbol{\rho}(x,y) \times \hat{\mathbf{v}}_t(x,y) ]p(x,y)dA    
\end{equation}
where $\boldsymbol{\rho}(x,y) = \begin{bmatrix} x & y \end{bmatrix}^T$ is a position vector with its origin at the \textcolor{blue3}{center} of pressure (CoP), as depicted in Fig. \ref{fig:contactArea}, while $\hat{\mathbf{v}}_t(x,y)$ denotes the unit velocity vector at coordinates $(x, y)$. The function $p(x,y)$ represents the pressure at $(x, y)$ and can be used to describe an arbitrary contact surface\textcolor{blue3}{;} see illustrations in Fig. \ref{fig:contactPressure}. For axisymmetric pressure distributions that exhibit circular contact pressure, the limit surface curve can be reduced to a two-dimensional curve within one quadrant. However, for more general pressure distributions, the limit surface takes the form of a three-dimensional surface.

\section{Planar friction modelling}\label{sec:full_model}

In this section, we introduce a distributed planar friction model that combines the principles of the LuGre model and the limit surface theory. \textcolor{blue}{The distributed model is extended to capture Elasto-Plastic behaviour to mitigate the drifting exhibited by the distributed LuGre model under oscillating loads.} Lastly, we show how to numerically compute the integrals for the distributed model.

\begin{figure}
    \centering
    \smallskip 
    \includegraphics[width=\columnwidth]{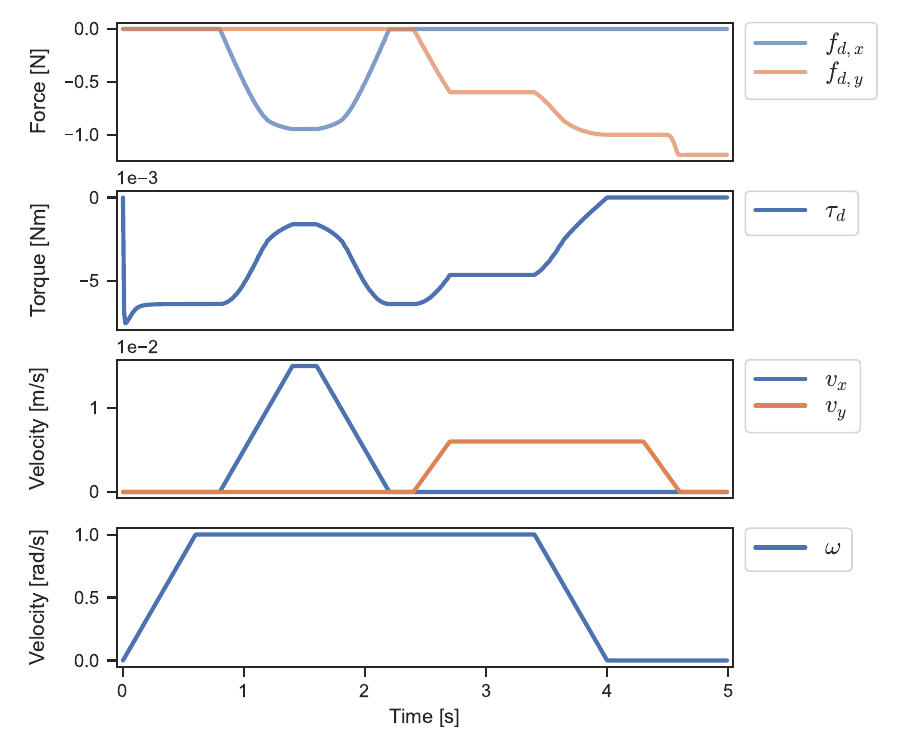}
    \caption{Simulation with predefined velocity profile for the distributed planar LuGre model. The simulation uses a circular contact surface with a radius of $10$ [mm] and a normal force of $1$ [N]. The velocity profile and friction force are at the CoP.}
    \label{fig:circle_dist_lugre}
    \vspace*{-0.5cm}
\end{figure}

\subsection{Distributed LuGre model}
The proposed model combines the LuGre model with the limit surface theory by extending the LuGre model \textcolor{blue}{with} 2D bristles and integrating them over the contact surface. A similar derivation in the context of tire friction has been done in \cite{velenis2002extension}, where the LuGre model was extended to 2D and integrated over one dimension. The integration over the contact surface allows for the representation of arbitrary contact pressures using the spatially varying function $p(x,y)$. We define the function $\mathcolor{blue}{\mathbf{l}_t}(\cdot)$ to describe the 2D LuGre friction, where the \textcolor{blue3}{area normalized} forces at $(x,y)$ are given by:
\begin{equation} \label{eq:2DLuGre}
\begin{split}
    \mathcolor{blue}{\mathbf{l}_t}&(\mathbf{v}_t(x,y), \mathbf{z}_t(x,y),  p(x,y)) = \\ & \left(\sigma_0  \mathbf{z}_t(x,y) + \sigma_1  \dot{\mathbf{z}}_t(x,y) + \sigma_2   \mathbf{v}_t(x,y) \right) p(x,y)
\end{split}    
\end{equation}
where $\mathbf{v}_t = \begin{bmatrix} v_x & v_y \end{bmatrix}^T$ represents the relative velocity between the surfaces (as shown in Fig. \ref{fig:contactArea}), and $\mathbf{z}_t = \begin{bmatrix} z_x & z_y \end{bmatrix}^T$ denotes the tangential bristle deflection. The rate of bristle deflection $\dot{\mathbf{z}}_t(x,y)$ is given by:

\begin{equation}\label{eq:z_dot_dist}
    \dot{\mathbf{z}}_t(x,y) = \mathbf{v}_t(x, y) -  \mathbf{z}_t(x,y) \frac{\sigma_0 ||\mathbf{v}_t(x,y)||}{g(||\mathbf{v}_t(x,y)||)}
\end{equation}
where $g(\cdot)$ is defined as in \eqref{eq:G_A}. The major difference in the 2D bristle deflection rate compared to equation \eqref{eq:dot_z} is the coupling between the $x$ and $y$ components due to the norm of the velocity $||\mathbf{v}_t(x,y)||$. This coupling ensures that the norm of the steady-state deflection is independent of the velocity direction. The steady-state deflection $\mathcolor{blue}{\mathbf{z}_t^\star}$ is given by:
\begin{equation}
    \mathcolor{blue}{\mathbf{z}_{t}^\star}(x,y) = \frac{ g(||\mathbf{v}_t(x,y)||) \mathbf{v}_t(x,y)}{\sigma_0 ||\mathbf{v}_t(x,y)||}
\end{equation}
and the local steady-state friction is represented as:
\begin{equation} \label{eq:2DLuGre_ss}
\begin{split}
    \mathcolor{blue}{\mathbf{l}_t^\star}&(\mathbf{v}_t(x,y), \mathcolor{blue}{\mathbf{z}_{t}^\star}(x,y), p(x,y)) = \\ & \left(\sigma_0  \mathcolor{blue}{\mathbf{z}_{t}^\star}(x,y) + \sigma_2  \mathbf{v}_t(x,y) \right) p(x,y)
\end{split}
\end{equation}   

To calculate the total tangential friction force, equation \eqref{eq:2DLuGre} is integrated over the contact surface:
\begin{equation} \label{eq:full_limit_lugre_ft}
   \mathcolor{blue}{\left[\begin{array}{c} f_{\textrm{d},x} \\ f_{\textrm{d}, y} \end{array}\right]} = -\int_A \mathcolor{blue}{\mathbf{l}_t}(\mathbf{v}_t(x,y), \mathbf{z}_t(x,y), p(x,y))  dA
\end{equation} 
where $\mathcolor{blue}{f_{\textrm{d},x}}$ and $\mathcolor{blue}{f_{\textrm{d}, y}}$ represent the tangential force at CoP \textcolor{blue}{for the distributed model}. The frictional torque at CoP can be calculated using:
\begin{equation}\label{eq:full_limit_lugre_tau}
    \mathcolor{blue}{\tau_\textrm{d}} = -\int_A \boldsymbol{\rho}(x,y) \times \mathcolor{blue}{\mathbf{l}_t}(\mathbf{v}_t(x,y), \mathbf{z}_t(x,y), p(x,y))  dA
\end{equation}
For the calculation of steady-state friction, equation \eqref{eq:2DLuGre_ss} is used instead of equation \eqref{eq:2DLuGre} in equations \eqref{eq:full_limit_lugre_ft} and \eqref{eq:full_limit_lugre_tau}.

\begin{figure}
    \centering
    \smallskip 
    \includegraphics[width=\columnwidth]{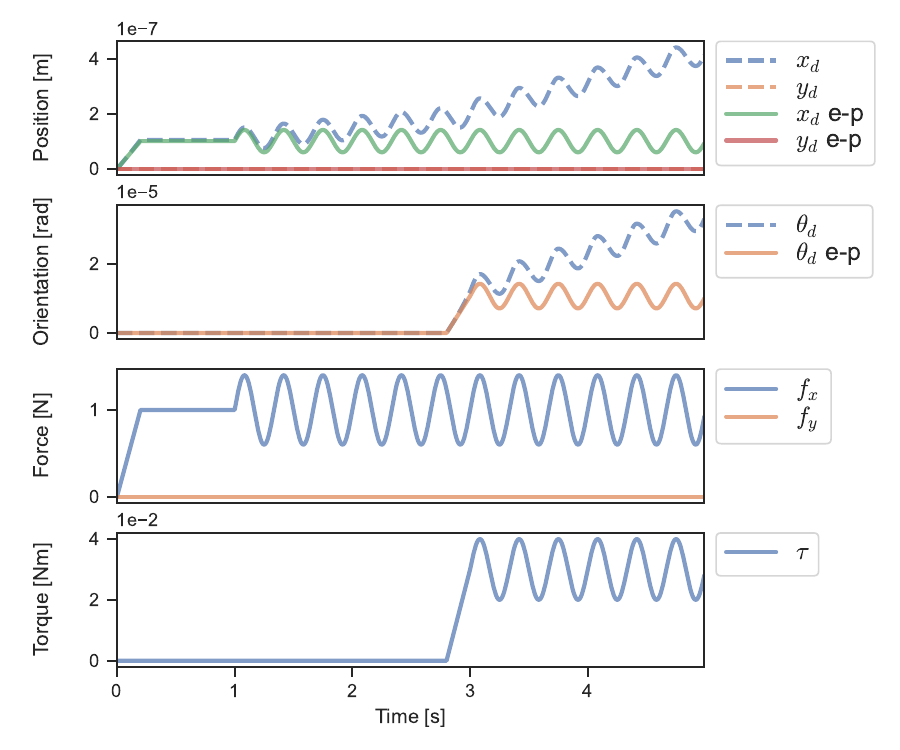}
    \caption{\textcolor{blue}{Object simulated with distributed} Lugre model drifting under \textcolor{blue}{imposed} oscillating tangential and torsional loads compared with the Elasto-Plastic (e-p) extension. The loads are under the threshold for initiating or sustaining slippage. Simulated with a $1$ [kg] disc with a radius of $0.05$ [m] and the settings specified in \textcolor{blue3}{T}able \ref{tab:sim_settings}.}
    \label{fig:2DLuGreDrift}
    \vspace*{-0.5cm}
\end{figure}

Throughout the paper, a running example is used to illustrate each section's contribution. An object with a circular contact area, $10$ [mm] radius, is placed on a flat surface and manifests $1$ [N] normal force. The object follows a preset velocity profile, and the resulting friction forces are simulated as shown in Fig. \ref{fig:circle_dist_lugre}. 
The velocity profile is designed such that the CoR moves inside or near the contact area to emphasize the coupling between the tangential and angular friction forces. If not explicitly stated otherwise, the examples and simulations use the settings provided in \textcolor{blue3}{T}able \ref{tab:sim_settings}, \textcolor{blue}{with the friction parameters $p=1$ from \textcolor{blue3}{T}able \ref{tab:parameters_num_cells}}.

{
\renewcommand{\arraystretch}{1.2}
\begin{table}[pbb]
\scriptsize
\centering
\caption{ Friction parameters \label{tab:parameters_num_cells}}
\resizebox{\columnwidth}{!}{%
\begin{tabular}{c|c|c|c|c|c|c|c|c}
\thickhline
\textbf{p} & $\sigma_0$ \textcolor{blue2}{[$1/\textrm{m}$]} & $\sigma_1$ \textcolor{blue2}{[$\textrm{s}/\textrm{m}$]}& $\sigma_2$ \textcolor{blue2}{[$\textrm{s}/\textrm{m}$]} & $\mu_c$ \textcolor{blue2}{[-]} & $\mu_s$ \textcolor{blue2}{[-]} & $\gamma$ \textcolor{blue2}{[-]} & $v_s$ \textcolor{blue2}{[$\textrm{m}/\textrm{s}$]} & $s_{\textrm{ba}}$ \textcolor{blue2}{[-]}\\ \thickhline
 0 & 1e6 & \textcolor{blue}{8e2} & 0 & 1 & 1 & 2 & 1e-3 & 0.9\\ \hline
 1 & 1e6 & \textcolor{blue}{8e2} & 0.2 & 1 & 1.2 & 2 & 1e-3 & 0.9\\ \hline
\end{tabular}
}
\end{table}
}

\subsection{Multi-dimensional Elasto-Plastic model}\label{sec:n_d_elasto_plastic}

\begin{figure}
    \centering
    \smallskip 
    \includegraphics[width=0.4\columnwidth]{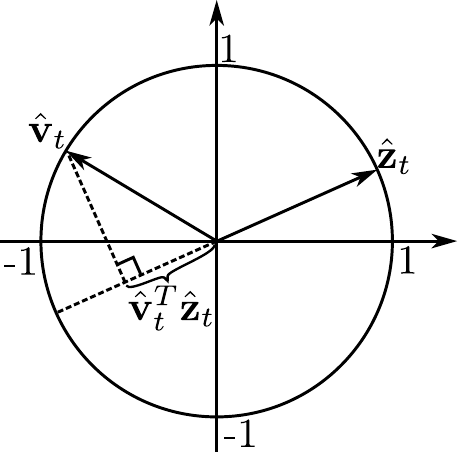}
    \caption{Projection of the unit velocity $\hat{\mathbf{v}}\mathcolor{blue3}{_t}$ on the unit bristle deflection $\hat{\mathbf{z}}\mathcolor{blue3}{_t}$.}
    \label{fig:unit_projection}
    \vspace*{-0.5cm}
\end{figure}

This section introduces an extension of the Elasto-Plastic model for multi-dimensional bristles, to address the drifting problem observed in the LuGre model for \textcolor{blue}{an object subjected to} oscillating tangential loads \textcolor{blue3}{(}see Fig. \ref{fig:2DLuGreDrift}\textcolor{blue3}{)}. The 1D Elasto-Plastic model has separate cases depending on if $z$ and $v$ have the same direction or not \eqref{eq:elasto_plastic_alpha_fun}. In this work, we generalize this concept to the planar case by a \textcolor{blue}{cosine similarity}. The Elasto-Plastic bristle deflection rate for the planar case is given by:
\begin{equation}\label{eq:elasto_plastic_ND}
    \dot{\mathbf{z}}_t(x,y) = \mathbf{v}_t(x, y) -  \mathbf{z}_t(x,y) \beta(\mathbf{z}_t, \mathbf{v}_t) \frac{\sigma_0 ||\mathbf{v}_t(x,y)||}{g(||\mathbf{v}_t(x,y)||)}
\end{equation}
where $\beta(\mathbf{z}_t, \mathbf{v}_t)$ is the extended version of $\alpha(z,v)$ in \eqref{eq:elasto_plastic_alpha_fun}. The key idea is to consider the alignment between the unit velocity vector $\hat{\mathbf{v}}_t$ and the unit bristle deflection vector $\hat{\mathbf{z}}_t$ to determine the elastic behaviour.
\textcolor{blue}{The cosine similarity} \textcolor{blue3}{(}see Fig. \ref{fig:unit_projection}\textcolor{blue3}{)} is used to calculate a ratio $\epsilon(\hat{\mathbf{z}}_t, \hat{\mathbf{v}}_t)$ that represents the alignment:
\begin{equation}
    \epsilon(\hat{\mathbf{z}}_t, \hat{\mathbf{v}}_t) = ( \hat{\mathbf{v}}_t^T \hat{\mathbf{z}}_t + 1)/2\mathcolor{blue3}{,}
\end{equation}
and the Elasto-Plastic variable $\beta(\mathbf{z}_t, \mathbf{v}_t)$ is calculated with: 
\begin{equation}\label{eq:beta_elasto_plastic}
    \beta(\mathbf{z}_t, \mathbf{v}_t) = \epsilon(\hat{\mathbf{z}}_t,  \hat{\mathbf{v}}_t) \bar{\beta}(\mathbf{z}_t)
\end{equation}
where the function $\bar{\beta}(\mathbf{z}_t)$ is similar to $\alpha(z, v)$ in \eqref{eq:elasto_plastic_alpha_fun}, \textcolor{blue3}{in particular} when $\sgn(z) = \sgn(v)$, but extended to use the norm instead. $\bar{\beta}(\mathbf{z}_t)$ is given by:  
\begin{equation}
    \bar{\beta}(\mathbf{z}_t) = \begin{cases} 0 & ||\mathbf{z}_t|| \leq z_\textrm{ba} \\ \alpha(||\mathbf{z}_t||) & z_\textrm{ba} \leq ||\mathbf{z}_t|| \leq z_\textrm{max}  \\ 1 & ||\mathbf{z}_t|| \leq z_\textrm{max}  \end{cases}
\end{equation}
where \textcolor{blue}{$\alpha(\cdot)$ is the function \textcolor{blue3}{of} \eqref{eq:alpha_z}}. The breakaway bristle deflection is defined as $z_\textrm{ba} = s_{\textrm{ba}} z_\textrm{max}$, where $s_{\textrm{ba}}$ is a scalar multiplier, and $z_\textrm{max} = ||\mathcolor{blue}{\mathbf{z}_{t}^\star}||$. For the 1D motion, this method is equivalent to the standard Elasto-Plastic method described in \ref{sec:elastPlastic}.

To validate the effectiveness of the Elasto-Plastic extension, simulations are conducted on a $1$ [kg] disc with a \textcolor{blue3}{$50$ [mm]} radius on a flat surface subjected to varying external loads. The loads are below the threshold \textcolor{blue}{to initiate or maintain slip}, see \textcolor{blue3}{Section} \ref{sec:results_drifing} for more details. The results in Fig. \ref{fig:2DLuGreDrift} demonstrate that the planar LuGre model without the Elasto-Plastic extension exhibits slight drift, while the model with the extension shows no drift, even when subjected to combined oscillating loads in the tangential and angular directions.

\subsection{Numerical approximation}\label{sec:numerical_approximation}
\begin{figure}
    \centering
    \smallskip 
    \includegraphics[width=\columnwidth]{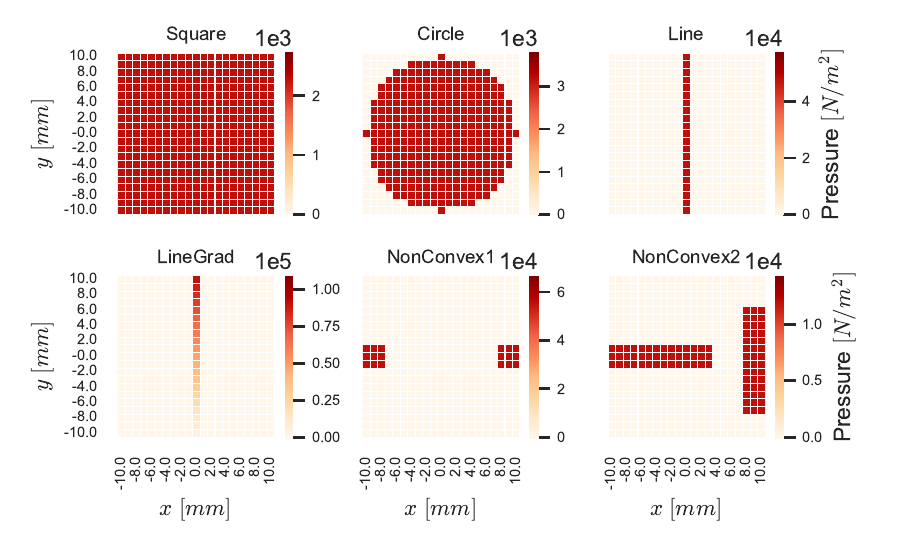}
    \caption{\textcolor{blue}{Examples of how different contact pressures are discretized. The contact area in the figure is discretized into $21 \times 21$ cells.}}
    \label{fig:ex_contact_pressure}
    \vspace*{-0.5cm}
\end{figure}

To solve the distributed model from equation \eqref{eq:full_limit_lugre_ft} and \eqref{eq:full_limit_lugre_tau}, the contact area is discretized into a grid of $N \times N$ cells, as shown in Fig. \ref{fig:ex_contact_pressure}. Each cells contains a 2D LuGre model, and the integrals in equations \eqref{eq:full_limit_lugre_ft} and \eqref{eq:full_limit_lugre_tau} can be discretized as follows:

\begin{align}
      \mathcolor{blue}{\left[\begin{array}{c} f_{\textrm{d},x} \\ f_{\textrm{d}, y} \end{array}\right]} &= -\sum\limits_{i_x = 1}^{N} \sum\limits_{i_y = 1}^{N} \mathcolor{blue}{\mathbf{l}_t}(\mathbf{v}_t(\cdot), \mathbf{z}_t(\cdot), p(\cdot)A)  \label{eq:full_limit_lugre_ft_d} \\
      \mathcolor{blue}{\tau_\textrm{d}} &= -\sum\limits_{i_x = 1}^{N} \sum\limits_{i_y = 1}^{N} \boldsymbol{\rho}(\cdot) \times \mathcolor{blue}{\mathbf{l}_t}(\mathbf{v}_t(\cdot), \mathbf{z}_t(\cdot), p(\cdot)A)  \label{eq:full_limit_lugre_tau_d}
\end{align}
where $A$ is the area of a cell.

{
\renewcommand{\arraystretch}{1.2}
\begin{table}[pbb]
\scriptsize
\centering
\caption{ Computational cost (distributed model)\label{tab:comp_time}}
\resizebox{\columnwidth}{!}{%
\begin{tabular}{c|c|c|c|c|c|c|c|c|c}
\thickhline
$N$ & 5 & 9  & 13 & 17 & 21 & 25 & 29 & 33 & 101\\ \thickhline
$\textrm{it}/\textrm{s}$ & 1.6e5 & 3.9e4 & 2.8e4 & 1.7e4 & 1.1e4 & 7.7e3 & 5.7e3 & 4.5e3 & 4.3e2\\ \hline
\end{tabular}
}
\end{table}
}

\begin{figure}
    \centering
    \smallskip 
    \includegraphics[width=\columnwidth]{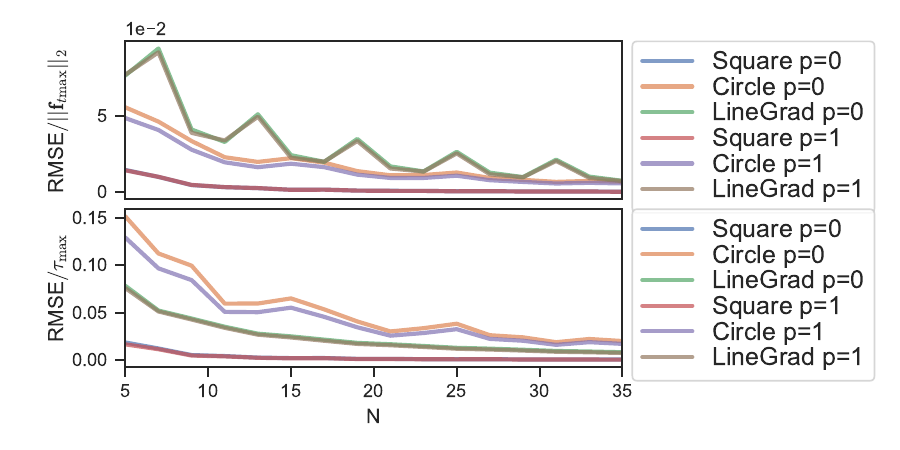}
    \caption{RMSE normalized with the maximum force over different numbers of cells $N \times N$ \textcolor{blue}{while following the velocity profile in Fig. \ref{fig:circle_dist_lugre}}. The results are for the planar distributed model, tested with different contact surfaces (see Fig. \ref{fig:ex_contact_pressure}) and different friction parameters (see \textcolor{blue3}{T}able \ref{tab:parameters_num_cells}). The top plot is for tangential friction and the bottom is for the torque. The velocity and force are evaluated at CoP.}
    \label{fig:num_cells}
    \vspace*{-0.5cm}
\end{figure}

The number of cells impacts the computational cost \textcolor{blue}{and the accuracy in relation to the size of the contact surface}. \textcolor{blue}{Fig. \ref{fig:num_cells} illustrates how the root mean square error (RMSE) normalized by the maximum force changes with the number of cells for different combinations of contact pressure distributions (see Fig. \ref{fig:ex_contact_pressure}) and friction parameters, as shown in \textcolor{blue3}{T}able \ref{tab:parameters_num_cells}}. The RMSE is calculated using a fixed velocity profile, as depicted in Fig. \ref{fig:circle_dist_lugre}. The ground truth is obtained using a $101 \times 101$ cell grid. \textcolor{blue}{The simulation is computed with a fixed time step of $\Delta t = 1\textrm{e}{-5}$ and} the average computational cost associated with different numbers of cells is summarized in \textcolor{blue3}{T}able \ref{tab:comp_time}, \textcolor{blue}{where $\textrm{it}/\textrm{s}$ stands for iterations per second, see \textcolor{blue3}{Section} \ref{sec:results} for the hardware used.}

\subsection{Bilinear steady-state approximation}\label{sec:bilinear}
\begin{figure}
    \centering
    \smallskip 
    \includegraphics[width=\columnwidth]{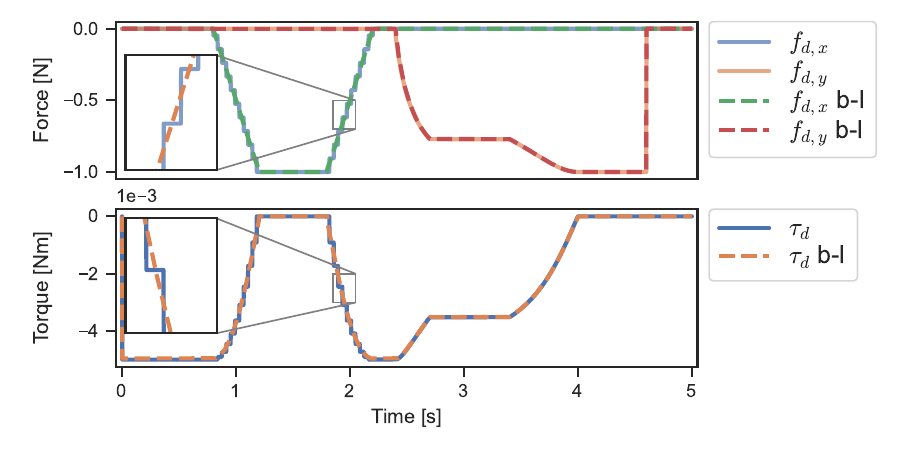}
    \caption{Discrete step behaviour of the steady-state friction due to the numerical approximation \eqref{eq:full_limit_lugre_ft_d} and \eqref{eq:full_limit_lugre_tau_d}. The results are simulated using a line contact following the velocity profile in \ref{fig:circle_dist_lugre}. The bilinear approximation (b-l) mitigates the step-like behaviour. Simulated with $p=0$ from \textcolor{blue3}{T}able \ref{tab:parameters_num_cells}.}
    \label{fig:bilinear}
    \vspace*{-0.5cm}
\end{figure}
The numerical approximation employed in equations \eqref{eq:full_limit_lugre_ft_d} and \eqref{eq:full_limit_lugre_tau_d} exhibits undesirable traits due to the discretization. \textcolor{blue}{The relative velocity of the surface can be represented as a pure rotation around the CoR}. In situations where the CoR moves within the contact area, discernible discontinuities manifest in the forces, as depicted in Fig. \ref{fig:bilinear}. The steady-state friction, essential for pre-computing the limit surface in \textcolor{blue3}{Section} \ref{sec:reduced_planar_friction}, is adversely affected by these discontinuities. \textcolor{blue}{Similar behaviour manifests for the non-steady state friction, however, bilinear interpolation for this case is less trivial to employ without significantly affecting the computational cost and is not considered in this paper.} 

\begin{figure}
    \centering
    \smallskip 
    \includegraphics[width=0.7\columnwidth]{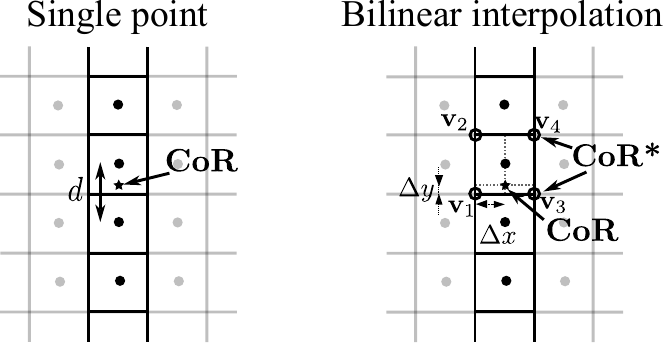}
    \caption{Illustrates the cells for a line contact. \textcolor{blue}{The friction contribution of each cell is calculated w.r.t. the \textcolor{blue3}{center} of that cell. On the left-hand side, the true relative velocity is used to evaluate the friction force. On the right-hand side, the friction force is estimated with bilinear interpolation between the forces resulting from a rotation around each corner of the cell that encloses the CoR.} The left-hand figure illustrates the distance $d$ that the CoR can travel before passing the midpoint of a cell.}
    \label{fig:bilinear_cor}
    \vspace*{-0.5cm}
\end{figure}

For a line contact, with the CoR traversing along its length, the tangential friction force exhibits increments upon surpassing the midpoint of a cell, as illustrated in Fig. \ref{fig:bilinear_cor}. \textcolor{blue}{The force increment is caused by the sudden change of velocity direction at the \textcolor{blue3}{center} of a cell when the CoR passes over it}. To circumvent this issue, a bilinear interpolation \cite{press2007numerical} method is employed \textcolor{blue}{for calculating steady-state friction}. First, the cell encompassing the CoR is identified, \textcolor{blue}{then the friction forces are calculated by applying the rotation around the temporary CoRs (CoR*) located at each corner of said cell. Let $\mathbf{v}_1$ to $\mathbf{v}_4$ be the velocity vectors that describe the rotation around each CoR*, the resulting friction forces are combined via bilinear interpolation:
\begin{equation}\label{eq:bilinear_f}
\begin{split}
        \mathbf{f}_{b}^\star (\mathbf{v}) = & (1-\Delta x)(1-\Delta y)\mathbf{f}_{d}^\star (\mathbf{v}_1) \\
         +&  (1-\Delta x)\Delta y\mathbf{f}_{d}^\star(\mathbf{v}_2) \\ 
         +& \Delta x(1-\Delta y)\mathbf{f}_{d}^\star(\mathbf{v}_3) \\ 
         +&  \Delta x \Delta y\mathbf{f}_{d}^\star(\mathbf{v}_4) \\ 
\end{split}
\end{equation}
where $\Delta x$ and $\Delta y$ are normalized distanced. $\mathbf{f}_{b}^\star (\mathbf{v}) $ is the estimate of the steady-state friction for the true CoR, as illustrated on the right-hand side of Fig. \ref{fig:bilinear_cor}}. A comparison between utilizing a single point and bilinear interpolation is shown in Fig. \ref{fig:bilinear}.

\section{Reduced planar friction \textcolor{blue3}{model}}\label{sec:reduced_planar_friction}
\begin{figure}
    \centering
    \begin{subfigure}[b]{0.59\columnwidth}
        \includegraphics[width=\columnwidth]{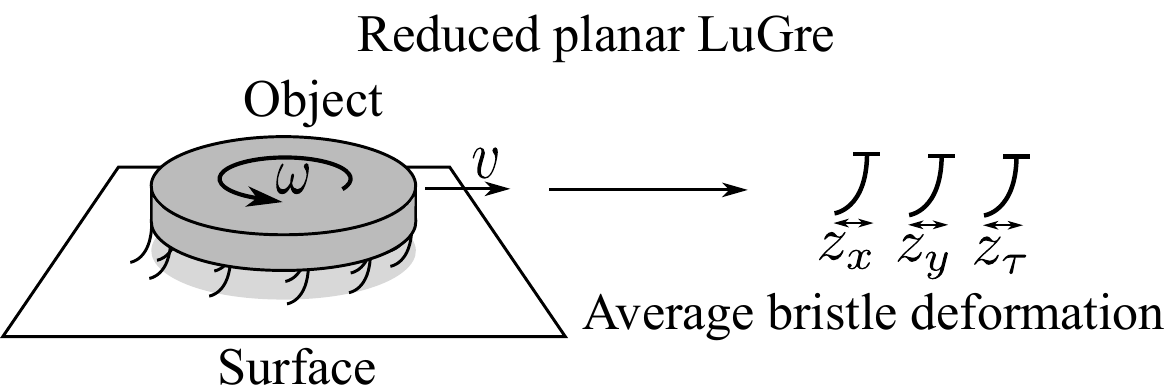}
        \caption{Bristle reduction}
    \end{subfigure}
    \begin{subfigure}[b]{0.39\columnwidth}
        \includegraphics[width=\columnwidth]{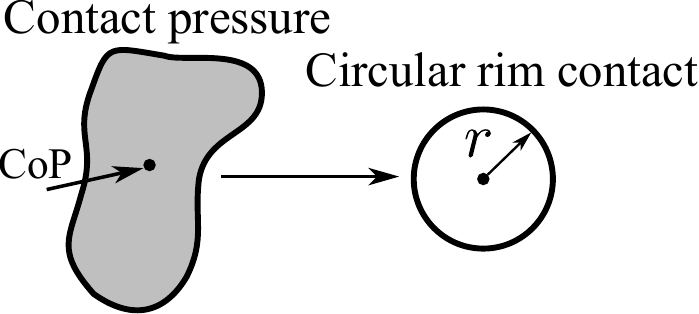}
        \caption{Circular rim contact}
        \label{fig:circular_rim_contact}
    \end{subfigure}
    \caption{The planar friction model reduced into a circular rim contact, which can be described by three bristles.}
    \label{fig:reduced_overview}
    \vspace*{-0.5cm}
\end{figure}

The primary concern with the distributed planar friction model lies in its computational complexity. As indicated in \textcolor{blue3}{T}able \ref{tab:comp_time}, the computational cost grows with the increasing number of cells employed to approximate the contact surface. In this section, we propose a novel approximation to the distributed planar LuGre model based on three bristles \textcolor{blue3}{(}see Fig. \ref{fig:reduced_overview}\textcolor{blue3}{)}. The three bristles are coupled with an ellipsoid approximation of the limit surface. The limitations of the ellipsoid approximation are discussed and we propose a \textcolor{blue}{method} based on a pre-computed limit surface to correct for the discrepancies. Finally, the pre-computation of the ellipsoid variables and the limit surface are discussed.  

\subsection{Ellipsoid approximation}\label{sec:ellipsoid_appox}

The distributed planar LuGre model is reduced to three bristles by approximating the contact surface as a circular rim contact, see figure \ref{fig:circular_rim_contact}. To achieve this, the coupling between the steady-state deflection of the three bristles is governed by an ellipsoid approximation of the limit surface.
The ellipsoid approximation with Coulomb friction has been employed in previous works \cite{shi2017dynamic}, \cite{xydas1999modeling}, and \cite{bicchi1993experimental}. The force wrench $\mathbf{f}_\textrm{e}$ is expressed as a function of the combined relative velocity \textcolor{blue}{twist} vector $\mathbf{v} = \begin{bmatrix} v_x &  v_y & \omega \end{bmatrix}^T$, which contains both the linear and angular velocity of a frame located at CoP. \textcolor{blue}{Defining $r$ as the radius of a corresponding circular rim contact, illustrated in Fig. \ref{fig:circular_rim_contact}, the twist velocity vector is scaled to $\mathbf{v}_S=\mathbf{S}\mathbf{v}$ where all elements have identical units and $\mathbf{S} = \textrm{diag}(\left[\begin{array}{ccc}     
    1 & 0 & r  
\end{array}\right])$. Under the ellipsoid approximation, the  mapping of $\mathbf{v}_S$ to the force wrench $\mathbf{f}_\textrm{e}$ for Coulomb friction is defined as follows:}
\begin{align}\label{eq:LS_vel}
    \mathbf{f}_\textrm{e} &= \mu_C \mathcolor{blue}{f_N} \frac{\mathcolor{blue}{\mathbf{S}} \mathbf{v}_S}{\mathcolor{blue}{||\mathbf{v}_S||}}
    \end{align}
\textcolor{blue}{where $\mathbf{S}$ scales the last element of $\mathbf{f}_e$ to be torque based on the radius $r$.} Having defined the force wrench in (\ref{eq:LS_vel}), few remarks are in order:
    \begin{itemize}
        \item The matrix  $\mathcolor{blue}{\mathbf{S}^2}$ determines the shape of the ellipsoid $ \frac{1}{(\mu_C \mathcolor{blue}{f_N})^2}\mathbf{f}_\textrm{e}^T\mathcolor{blue}{(\mathbf{S}^2)^{-1}} \mathbf{f}_\textrm{e}  = 1$.
        \item The radius $\mathcolor{blue}{r}$ can be pre-calculated for a surface as:
 \begin{equation}\label{eq:r_a}
    \mathcolor{blue}{r} = |\int_A [\boldsymbol{\rho}(x, y) \times \mathcolor{blue}{ \hat{\mathbf{v}}^\textrm{pr}_t}(x, y)]p_n(x,y) dA| 
\end{equation}
\textcolor{blue}{if the normalized pressure $p_n(x,y) = p(x, y)/f_N$ is constant}. $\mathcolor{blue}{ \hat{\mathbf{v}}^\textrm{pr}_t}(x, y)$ \textcolor{blue}{describes the relative unit velocity at $(x,y)$ for a pure rotation ($\mathcolor{blue}{\omega=1}$) around CoP.}
    \end{itemize}

The coupling between the steady-state deflection of the three bristles is modelled by replacing the Coulomb friction in \eqref{eq:LS_vel} with the 1D steady-state deflection, resulting in:
\begin{equation}
     \mathcolor{blue}{\mathbf{z}^{\star}} = \frac{ g(\mathcolor{blue}{||\mathbf{v}_S||}) \mathcolor{blue}{\mathbf{S}} \mathbf{v}_S } {\sigma_0 \mathcolor{blue}{||\mathbf{v}_S||}}
 \end{equation}
 By unravelling the expression, the bristle deflection rate can be determined as:
 \begin{equation}\label{eq:z_rate_ellipse}
     \dot{\mathbf{z}} = \mathcolor{blue}{\mathbf{S}} \mathbf{v}_S -  \mathbf{z}\frac{\sigma_0 \mathcolor{blue}{||\mathbf{v}_S||}}{ g(\mathcolor{blue}{||\mathbf{v}_S||})}
 \end{equation}
where $\mathbf{z} = \begin{bmatrix}
     z_x & z_y & z_\tau  
\end{bmatrix}^T$.

\begin{figure}
    \centering
    \smallskip 
    \includegraphics[width=\columnwidth]{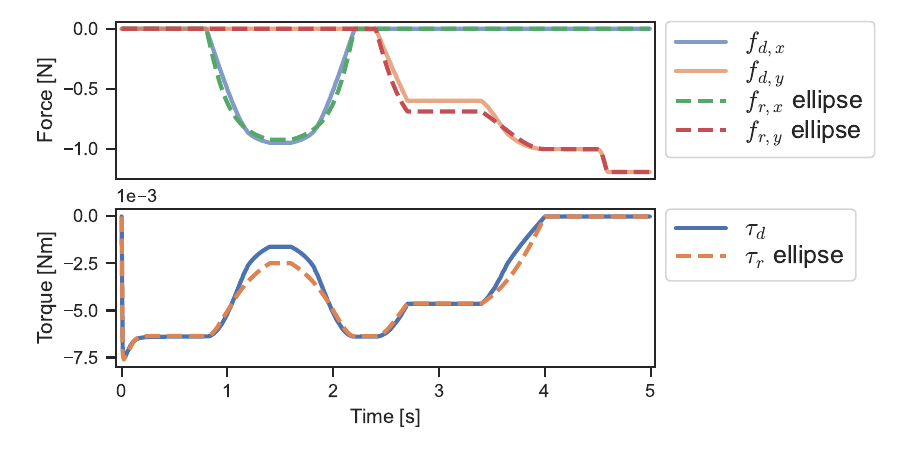}
    \caption{The reduced model with the ellipsoid approximation compared with the distributed model for the simulation in Fig. \ref{fig:circle_dist_lugre}.}
    \label{fig:ellipse_red_model}
    \vspace*{-0.5cm}
\end{figure}

The forces at COP are calculated by:
\begin{equation} \label{eq:red_get_f}
    \mathcolor{blue}{\mathbf{f}_{\textrm{r}}} = -(\sigma_0 \mathbf{z} + \sigma_1  \dot{\mathbf{z}} + \sigma_2 \mathbf{U} \mathbf{v}) f_N
\end{equation}
\textcolor{blue}{where $\mathbf{U} = \textrm{diag}(\begin{bmatrix}
1 & 1 & u \end{bmatrix})$ is a matrix for the viscous friction. If $p_n(x,y)$ is constant, then $u$ can be pre-calculated for a surface, as follows: 
\begin{equation}\label{eq:psi}
    u = \mathcolor{blue}{\int_A ||\boldsymbol{\rho}(x,y)||^2 p_n(x, y)  dA }
\end{equation}
which comes directly from the viscus friction of the distributed model under rotation around CoP:
\begin{equation}\label{eq:viscus_d}
    \mathcolor{blue}{\tau_{d, \nu} =\sigma_2 \omega f_N \int_A ||\boldsymbol{\rho}(x,y)||^2 p_n(x, y)  dA } 
\end{equation}
It should be noted that if $p_n(x, y)$ changes, then $u$ needs to be recalculated.}

The friction forces are simulated with the same example as in Fig. \ref{fig:circle_dist_lugre}, and the reduced planar LuGre model is compared with the distributed model \textcolor{blue3}{(}see Fig. \ref{fig:ellipse_red_model}\textcolor{blue3}{)}. It can be observed that the reduced model \textcolor{blue}{with ellipsoid approximation} does not exactly replicate the behaviour of the distributed model.

\subsection{Limitations with the ellipsoid approximation} \label{sec:ellipse_approximation}

\begin{figure}
    \centering
    \smallskip 
    \includegraphics[width=\columnwidth]{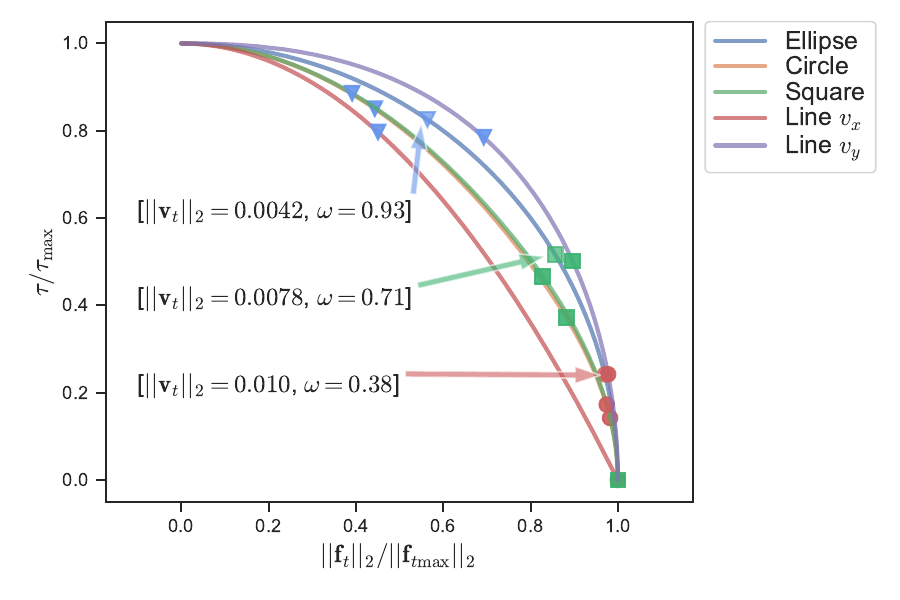}
    \caption{The limit curves for different contact surfaces compared with the ellipse approximation of a circle contact. The markers on each curve indicate the position of a specific velocity. In the case of the line contact, the tangential velocity $\mathbf{v}_t$ was tested along both the x-axis and y-axis, whereas for all other curves, only the x-axis was taken into consideration. $101 \times 101$ cells were used with $p=0$ from \textcolor{blue3}{T}able \ref{tab:parameters_num_cells}, i.e. Coulomb friction. }
    \label{fig:ellipse_p0}
    \vspace*{-0.5cm}
\end{figure}
\begin{figure}
    \centering
    \smallskip 
    \includegraphics[width=0.7\columnwidth, trim={0 0 0 1.5cm},clip]{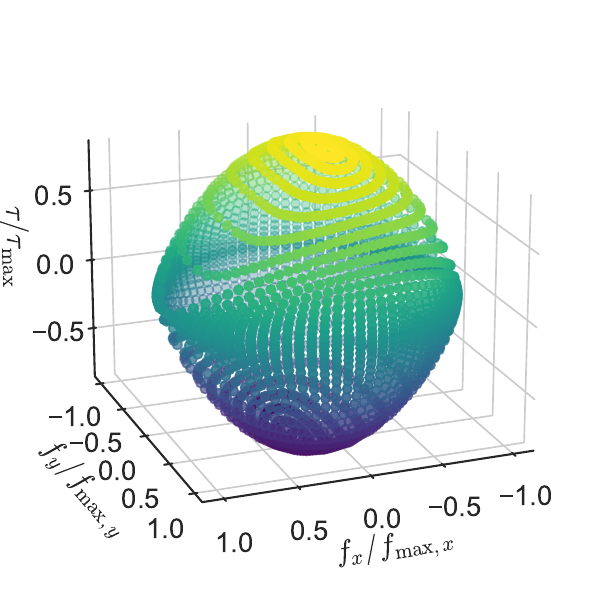}
    \caption{Normalized limit surface with Coulomb friction for a gradient line contact.}
    \label{fig:limit_surface}
    \vspace*{-0.5cm}
\end{figure}

The ellipsoid approximation of the limit surface is often considered a good approximation for two main reasons. First, it is computationally simple, which makes it practical for various applications. Second, the shape of the limit surface visually resembles an ellipsoid, as depicted in Fig. \ref{fig:ellipse_p0} and \ref{fig:limit_surface}. However, the shape of the limit surface can be deceiving, another crucial factor to consider is the correspondence between a given velocity and its location on the limit surface. Fig. \ref{fig:ellipse_p0} illustrates that the position on the limit surface associated with a particular velocity is highly correlated with the contact surface shape. Even for a circular shape, the ratio between torque and tangential force can vary significantly from the ellipsoid approximation. This is the main reason for the discrepancy observed in Fig. \ref{fig:ellipse_red_model} between the reduced model and the distributed model. 

\begin{figure}
    \centering
    \smallskip 
    \includegraphics[width=\columnwidth]{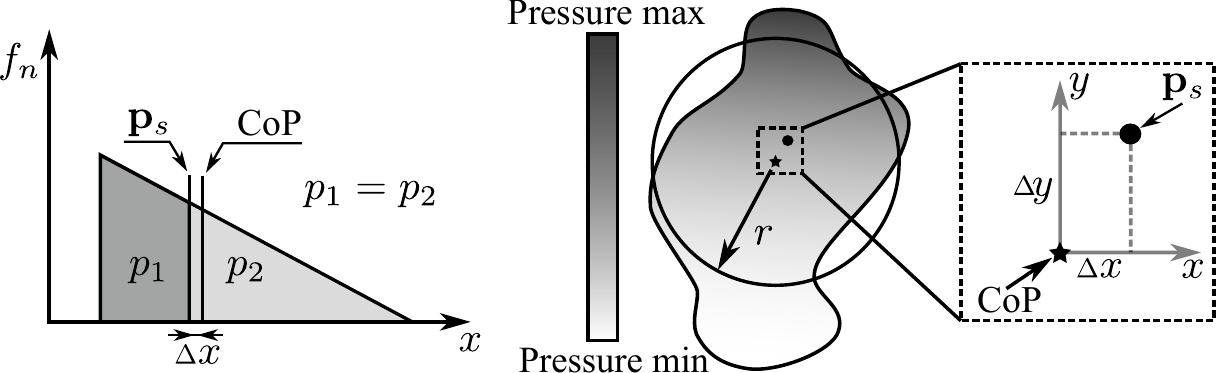}
    \caption{The left-hand side illustrates that CoP and $\mathbf{p}_s$ do not coincide for a gradient line contact. The $\mathbf{p}_s$ for a 1D surface has an equal amount of pressure on either side, $p_1 = p_2$, and a rotation around $\mathbf{p}_s$, therefore, generates zero tangential force \textcolor{blue}{under Coulomb friction}. CoP has zero moments due to the pressure, but a rotation around CoP can generate tangential forces. The right-hand side illustrates the concept for a 2D surface, where $\mathbf{p}_s$ is a rotation point with zero tangential forces.}
    \label{fig:skew_variables}
    \vspace*{-0.5cm}
\end{figure}

\textcolor{blue}{The ellipsoid approximation from \eqref{eq:LS_vel} will always express zero tangential force when the contact surface is subjected to a pure rotation around CoP. This is however not the case for the gradient line surface. Fig. \ref{fig:skew_variables} illustrates that there is a point $\mathbf{p}_s$ where a pure rotation around this point would generate zero tangential friction forces and that \textcolor{blue3}{$\mathbf{p}_s$} is not aligned with the CoP for the gradient line contact. The reason for this discrepancy is that the tangential friction force mainly depends on the pressure and the sign of the relative velocity at a patch on the contact surface, while CoP only depends on the distribution of the pressure.}

Based on these observations, it becomes evident that the ellipsoid approximation may be a less accurate approximation than previously suggested in the literature. While it still has many practical use cases due to its simplicity, it should be utilized with caution. In the next \textcolor{blue3}{Section}, \ref{sec:ls_scaling}, we will introduce a correction to the ellipsoid approximation method described in \textcolor{blue3}{Section} \ref{sec:ellipsoid_appox}.

\subsection{Limit surface scaling}\label{sec:ls_scaling}

In this section, we derive a scaled bristle deflection rate that follows a pre-computed limit surface which is normalized with the maximum force \textcolor{blue3}{and is defined by the function:} 
\begin{equation}\label{eq:h_pre}
    \mathbf{h}(\mathcolor{blue}{r}, \mathbf{v}) \triangleq \mathcolor{blue}{ \begin{bmatrix}
    \frac{f^\textrm{ls}_{x}(r, \mathbf{v})}{\max(f_x^\textrm{ls})} &  \frac{f^\textrm{ls}_{y}(r, \mathbf{v})}{\max(f_y^\textrm{ls})} & 
    \frac{\tau^\textrm{ls}(r, \mathbf{v})}{\max(\tau^\textrm{ls})}
\end{bmatrix}}^T
\end{equation}
\textcolor{blue}{where $-1 \leq \mathbf{h}_i(r, \mathbf{v}) \leq 1$ for each element $i$. The \textcolor{blue3}{function} $\mathbf{h}(\mathcolor{blue}{r}, \mathbf{v})$ for a gradient line contact is visualized in Fig. \ref{fig:limit_surface} and the calculation of $\mathbf{h}(\mathcolor{blue}{r}, \mathbf{v})$ is further discussed in \textcolor{blue3}{Section} \ref{sec:pre-calc}.}
The function $\mathbf{h}(\mathcolor{blue}{r}, \mathbf{v})$ is used to compute the corresponding steady-state bristle deflection for the limit surface as:
\begin{equation}\label{eq:esgn_LS}
    \mathcolor{blue}{ \mathbf{z}^\star} =  \frac{-\mathbf{S}\mathcolor{blue}{\mathbf{h}(r, \mathbf{v})} g(\mathcolor{blue}{||\mathbf{v}_S||})}{\sigma_0} 
\end{equation}
Unravelling \textcolor{blue}{the expression above} gives the bristle deflection rate:
 \begin{equation}\label{eq:z_rate_red}
     \dot{\mathbf{z}} =\mathcolor{blue}{ -\mathcolor{blue}{\mathbf{S}\mathbf{h}(r, \mathbf{v})} ||\mathbf{v}_S|| + \mathbf{z}  \frac{\sigma_0  ||\mathbf{v}_S||}{ g(\mathcolor{blue}{||\mathbf{v}_S||})} }
 \end{equation}
 where $\mathcolor{blue}{||\mathbf{v}_S||}$ is necessary for the bristle deflection rate.
 The Elasto-Plastic version can be formulated as:
  \begin{equation}\label{eq:z_rate_red_elasto}
     \dot{\mathbf{z}} = -\left[\mathcolor{blue}{\mathbf{S}\mathbf{h}(r, \mathbf{v})} + \mathbf{z} \beta(\mathbf{S}^{-1} \mathbf{z}, \mathcolor{blue}{-\mathbf{h}(r, \mathbf{v}) ||\mathbf{v}_S||}) \frac{\sigma_0}{ g(\mathcolor{blue}{||\mathbf{v}_S||})} \right] \mathcolor{blue}{||\mathbf{v}_S||}
 \end{equation}
where $\beta(\cdot, \cdot)$ is the function \textcolor{blue3}{defined} in \eqref{eq:beta_elasto_plastic}, here we use three-dimensional vectors as input. It is necessary to scale the input $\mathcolor{blue}{\mathbf{z}}$ to $\beta(\cdot, \cdot)$ with $\mathbf{S}^{-1}$ to ensure all elements have the same units. The forces at CoP are calculated by \eqref{eq:red_get_f} using the bristle deflection rate from \eqref{eq:z_rate_red_elasto}. 

\begin{figure}
    \centering
    \smallskip 
    \includegraphics[width=\columnwidth]{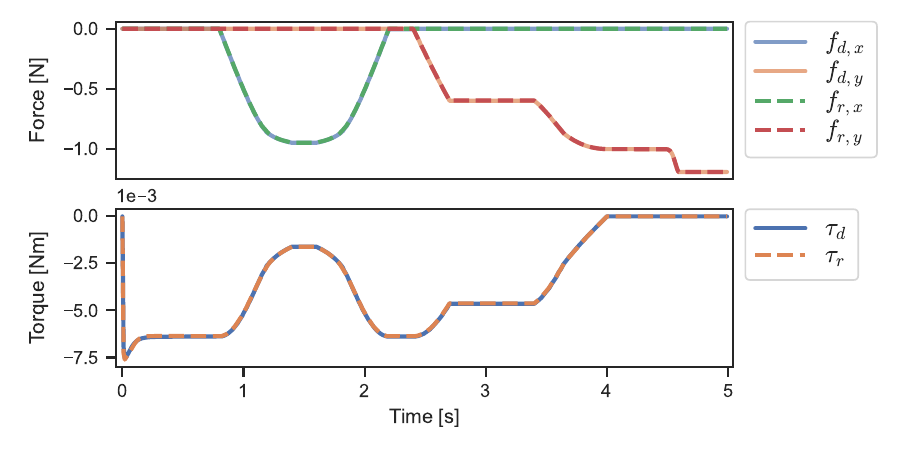}
    \caption{Simulated forces from the reduced friction model with the limit surface correction, along with the forces from the distributed model. See Fig. \ref{fig:circle_dist_lugre} for simulation details.}
    \label{fig:force_vel}
    \vspace*{-0.5cm}
\end{figure}

Fig. \ref{fig:force_vel} compares the reduced model with the limit surface correction to the planar distributed model. It demonstrates that the limit surface correction allows the model to closely match the behaviour of the distributed model. 
The subsequent section explains the pre-computation in detail.
 
\subsection{Pre-computation of limit surface}\label{sec:pre-calc}

\begin{figure}
    \centering
    \smallskip 
    \includegraphics[width=\columnwidth]{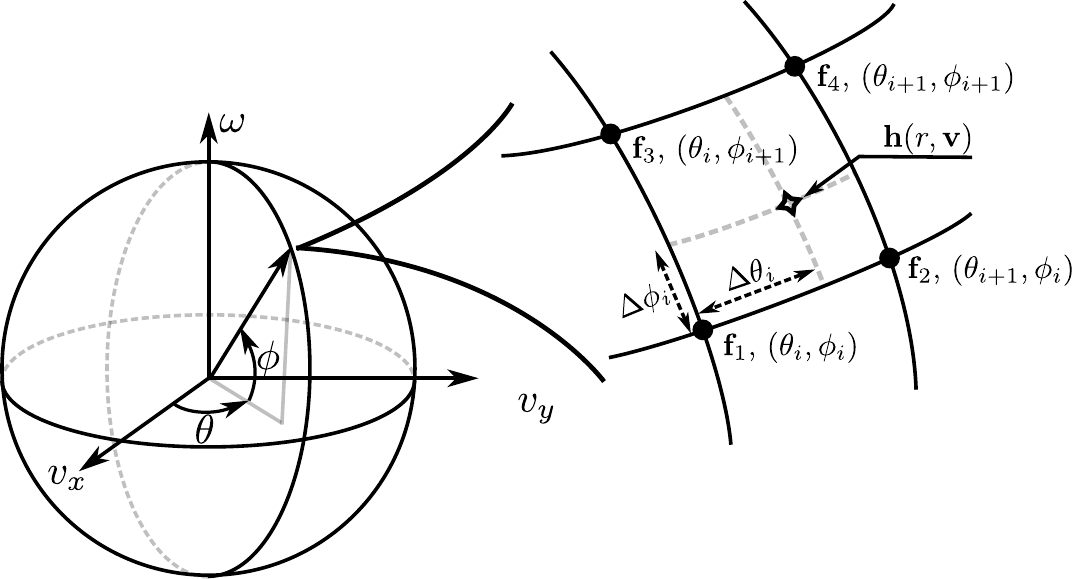}
    \caption{Illustrates the spherical coordinates used to sample the limit surface and the bilinear interpolation of a cell on the pre-computed limit surface.}
    \label{fig:spherical_coord}
    \vspace*{-0.5cm}
\end{figure}

\textcolor{blue}{In the preceding \textcolor{blue3}{Section} \ref{sec:ls_scaling}, we used the normalized limit surface function $\mathbf{h}(\mathcolor{blue}{r}, \mathbf{v})$ \textcolor{blue3}{(}see Fig. \ref{fig:limit_surface}\textcolor{blue3}{)} to calculate the scaled bristle deflection. Here, we introduce a method for pre-computing $\mathbf{h}(\mathcolor{blue}{r}, \mathbf{v})$ as a hash map with bilinear interpolation. To pre-compute $\mathbf{h}(\mathcolor{blue}{r}, \mathbf{v})$, velocity points $\mathbf{v}$ are sampled from spherical coordinates $0 \leq \theta \leq 2\pi$ and $0 \leq \phi \leq \pi/2$ \textcolor{blue3}{(}see Fig. \ref{fig:spherical_coord}\textcolor{blue3}{)} where $\theta$ represents the direction of the tangential velocity and $\phi$ represents the ratio between sliding and spinning motion. The angles $\theta $ and $\phi$ are sampled uniformly, however, the majority of the critical dynamics of the limit surface occur when the CoR lies within the contact area. Therefore, the tangential velocities are scaled with $\mathcolor{blue}{\mathcolor{blue}{r}'}$, where $\mathcolor{blue}{r'}$ is the \textcolor{blue3}{radius} 
of the surface used to pre-calculate $\mathcolor{blue}{\mathbf{h}(r, \mathbf{v})}$, to give a more even distribution of sampled forces. The sampled velocity vector is calculated as:
\begin{equation}
    \left[\begin{array}{c}
         v_x(\theta, \phi)  \\
         v_y(\theta, \phi) \\
         \omega(\phi)
    \end{array}\right] = \left[\begin{array}{c}
         r' \cos\theta \sin{\phi}  \\
         r' \sin\theta \sin{\phi} \\
         \cos{\phi}
    \end{array}\right]
\end{equation}}

Four neighbouring samples make up a cell \textcolor{blue3}{(}see Fig. \ref{fig:spherical_coord}\textcolor{blue3}{)}. As we assume equal friction properties in all directions, it suffices to sample only half of the sphere. Let $\mathcolor{blue}{N_{\textrm{ls}}}$ represent the number of cells between $0$ to $\pi/2$ [rad], and each cell on the sphere can be indexed with a unique hash map index:
\begin{equation}\label{eq:index_ls}
    \mathcolor{blue}{i_c} = \mathcolor{blue}{N_{\textrm{ls}}} \mathcolor{blue}{i_\theta} + \mathcolor{blue}{i_\phi}
\end{equation}
where $\mathcolor{blue}{i_\theta}$ and $\mathcolor{blue}{i_\phi}$ are the indices along $\theta$ and $\phi$, respectively. \textcolor{blue}{To compute the normalized force wrench $\mathbf{f}_\textrm{c}$ corresponding to a corner of a cell}, we utilize the steady-state interpolation method described in Section \ref{sec:bilinear} with the parameters $p=0$ from \textcolor{blue3}{T}able \ref{tab:parameters_num_cells} \textcolor{blue}{to numerically approximate the limit surface}, and the force wrench is normalised with the corresponding maximum force, see \eqref{eq:h_pre}.

To retrieve the normalized force wrench from $\mathbf{h}(\mathcolor{blue}{r}, \mathbf{v})$, the angle $\theta$ can be retrieved from a given velocity vector $\mathbf{v}$ by:
 \begin{equation}
     \theta = \begin{cases}
         \textrm{atan2}(v_y,  v_x) & \omega \geq 0 \\
         \textrm{atan2}(v_y,  v_x) + \pi & \omega < 0
     \end{cases}
 \end{equation}
 where $\theta$ always belongs to $0 \leq \theta < 2\pi$. For $\phi$, only the top half of the limit surface is considered: 
 \begin{equation}\label{eq:ls_shift}
     \phi = \textrm{atan2}(\mathcolor{blue}{r} |\omega|,||\mathbf{v}_t||)
 \end{equation}
where $\mathcolor{blue}{r}$ compensates for the stretched velocity sphere and allows the contact surface to be re-sized without re-calculation of $\mathbf{h}(\mathcolor{blue}{r}, \mathbf{v})$. 

The hash map index can be retrieved by calculating the corresponding indices of the angles: 
\begin{equation}
    \mathcolor{blue}{i_\theta} =\textrm{floor}(\frac{2 \theta \mathcolor{blue}{N_{\textrm{ls}}}}{\pi})
\end{equation}
\begin{equation}
    \mathcolor{blue}{i_\phi} = \textrm{floor}(\frac{2 \phi \mathcolor{blue}{N_{\textrm{ls}}}}{\pi})
\end{equation}
and then the hash map key is calculated using \eqref{eq:index_ls}. Within each cell, we employ bilinear interpolation, taking advantage of the residuals obtained from the flooring process:
\begin{equation} \label{eq:delta_theta}
    \Delta \mathcolor{blue}{i_\theta} = \frac{2 \theta \mathcolor{blue}{N_{\textrm{ls}}}}{\pi} - \mathcolor{blue}{i_\theta}
\end{equation}
\begin{equation}\label{eq:delta_phi}
    \Delta\mathcolor{blue}{i_\phi} =  \frac{2 \phi \mathcolor{blue}{N_{\textrm{ls}}}}{\pi} - \mathcolor{blue}{i_\phi}
\end{equation}
which range between 0 and 1 \textcolor{blue3}{(}see Fig. \ref{fig:spherical_coord}\textcolor{blue3}{)}. The residuals are used to interpolate between the \textcolor{blue}{forces $\mathcolor{blue}{\mathbf{f}_{\textrm{c}, 1}(i_c)}$ to $\mathcolor{blue}{\mathbf{f}_{\textrm{c}, 4}(i_c)}$ corresponding to the corners of a cell}, yielding the output: 
\begin{equation}
\begin{split}
        \mathbf{h}(\mathcolor{blue}{r}, \mathbf{v}) =&\mathcolor{blue}{\sgn_p(\omega)}((1 -  \Delta i_\theta)(1 - \Delta i_\phi)\mathbf{f}_{\textrm{c}, 1}(i_c) \\
         +& \Delta i_\theta(1 - \Delta i_\phi)\mathbf{f}_{\textrm{c}, 2}(i_c) \\ 
         +&(1 -  \Delta i_\theta)\Delta i_\phi\mathbf{f}_{\textrm{c}, 3}(i_c) \\ 
         +& \Delta i_\theta\Delta i_\phi\mathbf{f}_{\textrm{c}, 4}(i_c)) \\ 
\end{split}
\end{equation}
where
 \begin{equation}
 \mathcolor{blue}{
     \sgn_p(\omega) = \begin{cases}
         1 & \omega \geq 0 \\
         -1 & \omega  < 0
     \end{cases}}
 \end{equation}

\section{Results}\label{sec:results}

\begin{figure}
    \centering
    \smallskip 
    \includegraphics[width=\columnwidth]{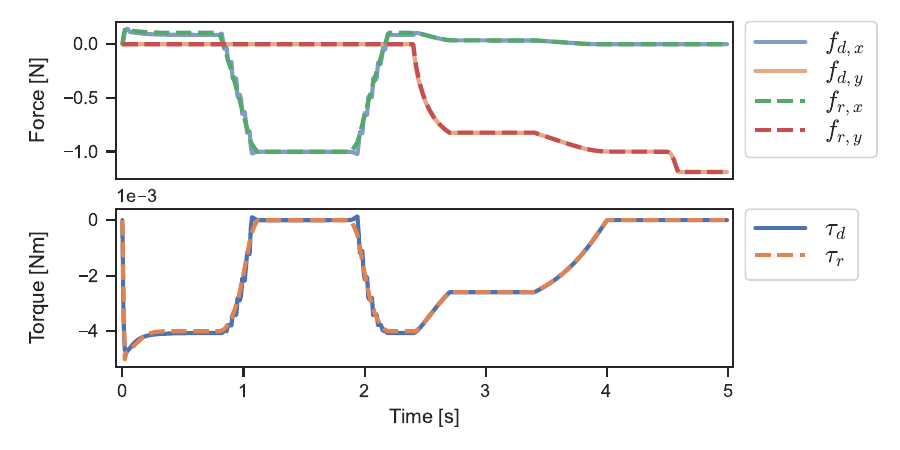}
    \caption{Forces from a gradient line contact following the velocity profile in Fig. \ref{fig:circle_dist_lugre}. The figure shows both the distributed model and the reduced model.}
    \label{fig:force_vel_line_grad}
    \vspace*{-0.5cm}
\end{figure}

\textcolor{blue}{This section expands on the results presented throughout the paper and presents new experiments to validate the proposed friction models}. \textcolor{blue3}{In \textcolor{blue3}{Section} \ref{sec:VA}}, the reduced model is compared to the planar distributed friction model using a fixed velocity profile. \textcolor{blue}{Then, the reduced model with and without the limit surface correction is compared.} \textcolor{blue}{In section \ref{sec:results_drifing}, the planar Elasto-Plastic extension and the drifting behaviours under oscillating loads are demonstrated. Moreover in \ref{sec:in_hand}, the friction models are simulated using an in-hand stick-slip experiment to assess their practical applicability. Finally, in \textcolor{blue3}{Section} \ref{sec:pre_comp_run_time}, the computational cost is assessed and the pre-computation time is analyzed.}
\begin{figure}
    \centering
    \smallskip 
    \includegraphics[width=\columnwidth]{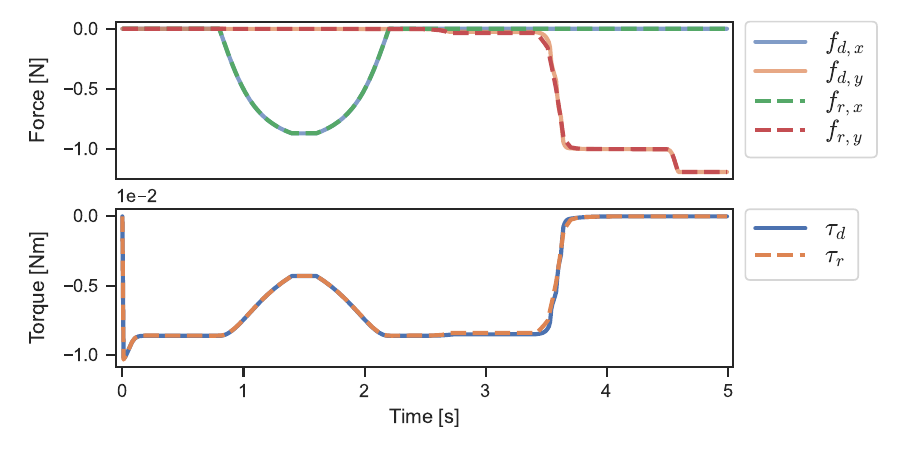}
    \caption{\textcolor{blue}{Forces from a gradient non-convex 1 contact following the velocity profile in Fig. \ref{fig:circle_dist_lugre}. The figure shows both the distributed model and the reduced model.}}
    \label{fig:force_vel_nonConvex1}
    \vspace*{-0.5cm}
\end{figure}

All experiments are conducted on an Intel i7-1185G7 processor \textcolor{blue}{with access to 32 GB of RAM}. Unless explicitly stated otherwise, the experiments adopt the settings in \textcolor{blue3}{T}able \ref{tab:sim_settings}. \textcolor{blue}{The LSODA solver is accessed through SciPy \cite{2020SciPy-NMeth}.} To ensure a fair comparison, the friction models are implemented in C++ and compiled as a Python module using pybind11 \cite{pybind11}.

{
\renewcommand{\arraystretch}{1.2}
\begin{table}[pbb]
\scriptsize
\centering
\caption{ Simulation settings \label{tab:sim_settings}}
\begin{tabular}{c|c}
\hline
Parameters (\textcolor{blue3}{T}able \ref{tab:parameters_num_cells}) & $p=1$\\ \hline
\textcolor{blue}{Solver} & \textcolor{blue}{LSODA} \\ \hline
\textcolor{blue}{Absolute tolerance} & $\mathcolor{blue}{1\textrm{e}{-8}}$ \\ \hline
\textcolor{blue}{Relative tolerance} & $\mathcolor{blue}{1\textrm{e}{-6}}$ \\ \hline
\textcolor{blue}{Max time step} & $\mathcolor{blue}{1\textrm{e}{-3}}$ \textcolor{blue2}{[s]} \\ \hline
Contact area & $0.02 \times 0.02$  \textcolor{blue2}{[$\textrm{m}^2$]}\\ \hline
Number of cells & $21 \times 21$\\ \hline
Pre-computation resolution & $\mathcolor{blue}{N_{\textrm{ls}}} = 20$\\ \hline
\end{tabular}

\end{table}
}

\subsection{Reduced vs planar distributed model}\label{sec:VA}

\begin{figure}
    \centering
    \smallskip 
    \includegraphics[width=\columnwidth]{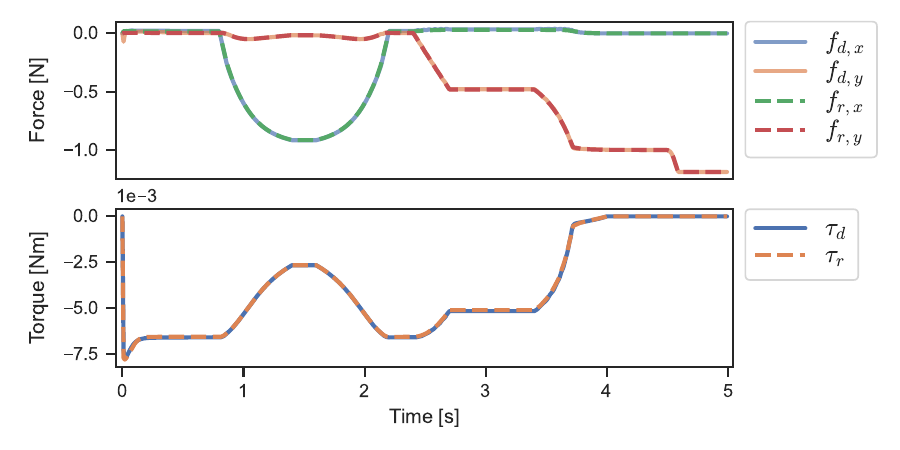}
    \caption{\textcolor{blue}{Forces from a gradient non-convex 2 contact following the velocity profile in Fig. \ref{fig:circle_dist_lugre}. The figure shows both the distributed model and the reduced model.}}
    \label{fig:force_vel_nonConvex2}
    \vspace*{-0.5cm}
\end{figure}
\begin{figure}
    \centering
    \smallskip 
    \includegraphics[width=\columnwidth]{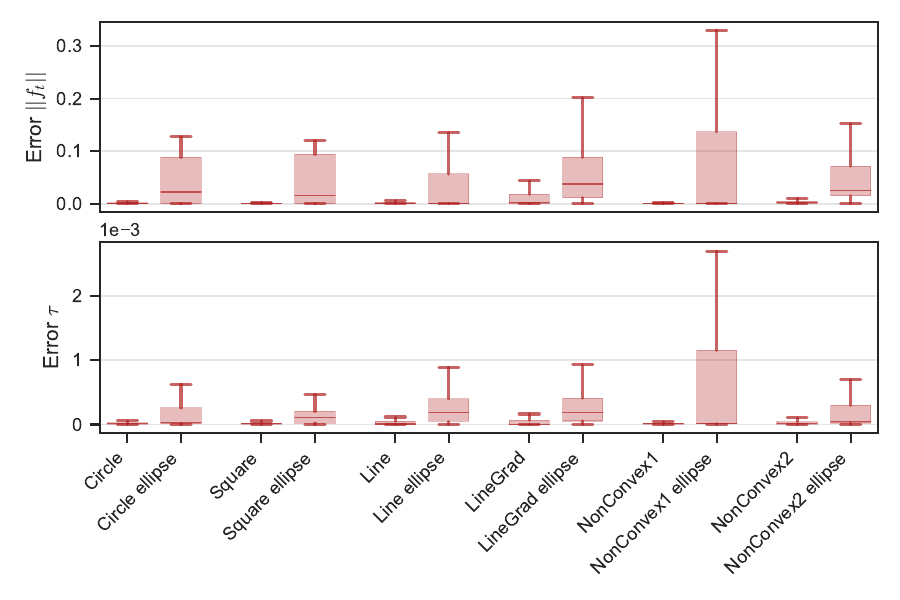}
    \caption{\textcolor{blue}{Boxplot comparing the error in forces between the reduced model with limit surface correction and the model with ellipsoid approximation. Both models follow the same fixed velocity, see fig \ref{fig:circle_dist_lugre}. The error is calculated by taking the difference to the distributed model. The boxplot represents the $\textrm{Q1}-1.5\textrm{IQR}$, Q1, median, Q3 and $\textrm{Q3}+1.5\textrm{IQR}$. }}
    \label{fig:error_bars} 
    \vspace*{-0.5cm}
\end{figure}

In the previous \textcolor{blue3}{Section} \ref{sec:reduced_planar_friction}, the reduced model and planar distributed model were compared in the context of a circular contact surface with a fixed velocity profile, \textcolor{blue3}{(see} Fig. \ref{fig:circle_dist_lugre}\textcolor{blue3}{)}. In this section, the same experiment is conducted but with a gradient line contact, and the resulting friction forces are depicted in Fig. \ref{fig:force_vel_line_grad}. It is evident that the friction forces and their transitions exhibit significant differences compared to those observed with a circular contact. Furthermore, it should be noted that in regions where pure rotation around the CoP occurs, the tangential friction force is non-zero. \textcolor{blue}{The reduced model incorporates this with the pre-computed limit surface.} Lastly, it is worth observing that the reduced model does not exhibit the \textcolor{blue}{effect} of discrete force steps as seen in the distributed model for linear shapes. This advantage arises due to the utilization of bilinear interpolation during the pre-computation stage.

\textcolor{blue}{The simulations in \textcolor{blue3}{Figs.} \ref{fig:force_vel} \textcolor{blue3}{and \ref{fig:force_vel_line_grad} -  \ref{fig:force_vel_nonConvex2}} all follow the same exact velocity profile presented in Fig. \ref{fig:circle_dist_lugre}. However, the resulting friction forces differ significantly depending on the pressure distribution. The two friction forces from the two non-convex surfaces from Fig. \ref{fig:ex_contact_pressure} are presented in Fig. \ref{fig:force_vel_nonConvex1} and \ref{fig:force_vel_nonConvex2}. The results show that the reduced friction model performs well even with challenging contact pressure distributions.}

\textcolor{blue}{In Fig. \ref{fig:error_bars}, the reduced model with the limit surface correction is compared to the reduced model with just the ellipsoid approximation. The experiment used the distributed model as ground truth. For all surfaces tested, the limit surface correction significantly outperforms the ellipsoid approximation. \textcolor{blue3}{Notably}, the contact surface with linear gradient pressure has a significantly higher error compared to the other surfaces with the limit surface correction. The authors attribute a large part of the error to the stepping effect of the distributed model for that surface \textcolor{blue3}{(}see Fig. \ref{fig:force_vel_line_grad}\textcolor{blue3}{)}.}

\begin{figure}
    \centering
    \smallskip 
    \includegraphics[width=\columnwidth]{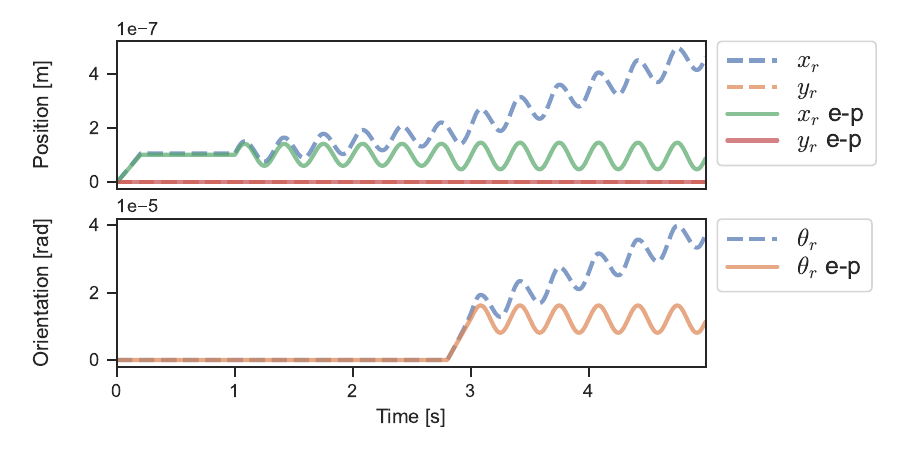}
    \caption{The reduced planar LuGre model exhibits drift when the oscillating tangential loads from Fig. \ref{fig:2DLuGreDrift} are applied, and the Elasto-Plastic (e-p) extension of the reduced model mitigates the drifting. Simulated with a 1 [kg] disc with a $0.05$ [m] radius.}
    \label{fig:2DLuGreDrift_reduced}
    \vspace*{-0.5cm}
\end{figure}
\begin{figure}
    \centering
    \smallskip 
    \includegraphics[width=\columnwidth]{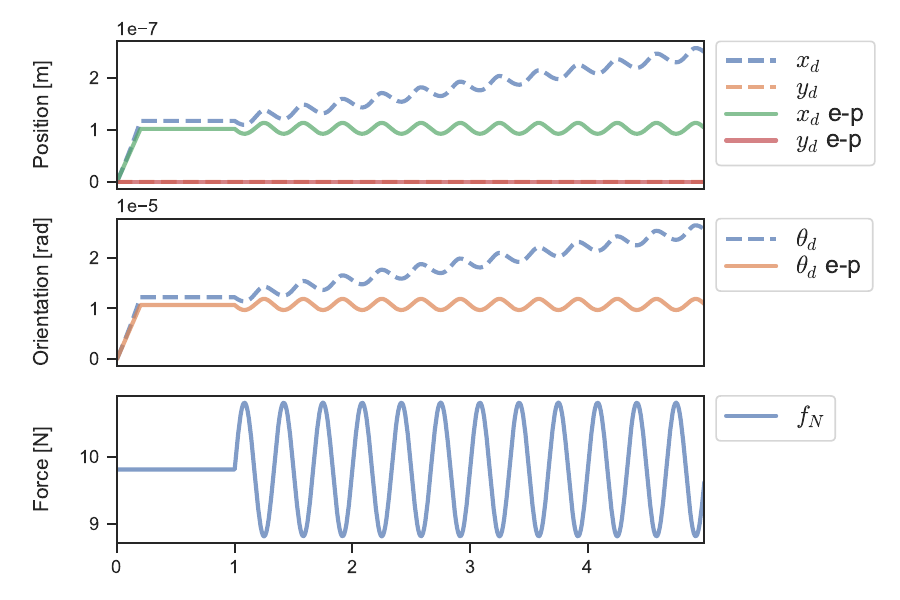}
    \caption{The distributed planar LuGre model exhibits drift when an oscillating normal load is applied together with a constant tangential load, and the Elasto-Plastic (e-p) extension mitigates the drifting, although oscillations in displacements still persist. Simulated with a 1 [kg] disc with a $0.05$ [m] radius. }
    \label{fig:2DLuGreDriftNormal}
    \vspace*{-0.5cm}
\end{figure}

\subsection{Drifting with oscillating tangential and normal load}\label{sec:results_drifing}

In this section, the planar Elasto-Plastic extension is tested on both planar friction models under oscillating loads. The simulation is conducted with a 1 kg disc having a radius of $0.05$ [m] on a flat surface, and the pose of the disc is tracked. In the first experiment, a tangential load equivalent to 1/12 of the force required to initiate slip is applied. After one second, an oscillation is introduced. At three seconds, an oscillating torque of approximately 1/6 of the torque required for slippage is applied \textcolor{blue3}{(}see Fig. \ref{fig:2DLuGreDrift}\textcolor{blue3}{)}. For the planar distributed model, the Elasto-Plastic extension mitigates the drifting observed in the LuGre model. This improvement is similar to the one-dimensional case described in \cite{dupont2000elasto}. The same experiment is conducted for the reduced model, as shown in Fig. \ref{fig:2DLuGreDrift_reduced}. Similarly, drifting occurs when the Elasto-Plastic extension is not utilized, but when it is employed, no drifting is observed. Here, there is a slight increase in the amplitude of the displacement oscillations when the torque is added.

\begin{figure}
    \centering
    \smallskip 
    \includegraphics[width=\columnwidth]{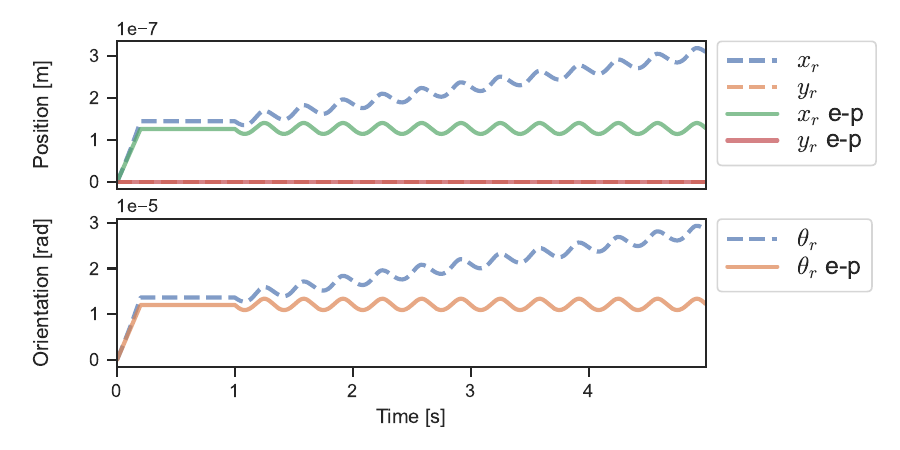}
    \caption{The reduced planar LuGre model also exhibits drift when an oscillating normal load is applied together with a constant tangential load. Similarly to the distributed model (Fig. \ref{fig:2DLuGreDriftNormal}), the Elasto-Plastic (e-p) extension mitigates the drifting. Simulated with a 1 [kg] disc with a $0.05$ [m] radius.}
    \label{fig:2DLuGreDrift_reduced_normal}
    \vspace*{-0.5cm}
\end{figure}

Furthermore, drifting in displacement occurs when an oscillating normal load is applied with a fixed tangential load, as depicted in Fig. \ref{fig:2DLuGreDriftNormal}. The same disc is used, and a constant force of $f_x = 1$ $[\textrm{N}]$ and $f_\tau = 0.03$ $[\textrm{Nm}]$ is applied. The normal load oscillates with an amplitude of $1$ $[\textrm{N}]$. From the results in Fig. \ref{fig:2DLuGreDriftNormal}, it can be observed that the Elasto-Plastic extension mitigates the drifting. However, the oscillations in displacement persist. Similar trends are observed in the results obtained from the reduced model, as shown in Fig. \ref{fig:2DLuGreDrift_reduced_normal}.

\subsection{In-hand slip-stick}\label{sec:in_hand}
\begin{figure}
    \centering
    \begin{subfigure}[b]{0.398\columnwidth}
        \includegraphics[width=\columnwidth]{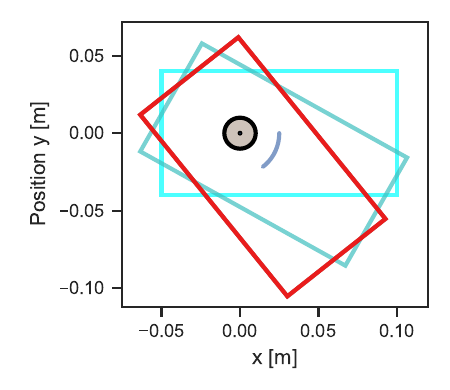}
        \caption{Case 1: CoM $25$ [mm] to CoA.}
        \label{fig:slip_stick_a}
    \end{subfigure}
    \begin{subfigure}[b]{0.589\columnwidth}
        \includegraphics[width=\columnwidth]{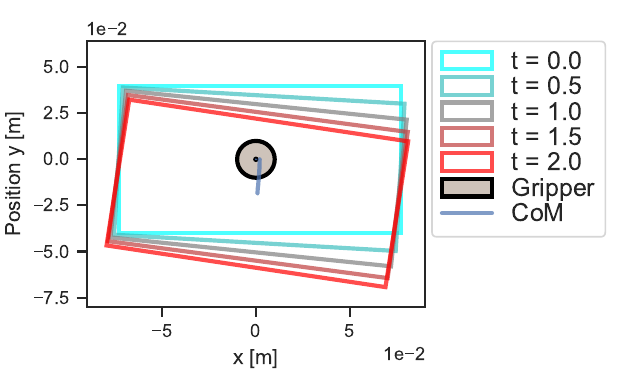}
        \caption{Case 2: CoM $2$ [mm] to CoA.}
        \label{fig:slip_stick_b}
    \end{subfigure}
    \caption{Snapshots from simulation using the sinusoidal normal force, the snapshots correspond to simulations in Fig. \ref{fig:slip_stick_circular_com} (left) and \ref{fig:slip_stick} (right).} 
    \label{fig:slip_stick_view}        
    \vspace*{-0.5cm}
\end{figure}
\begin{figure}
    \centering
    \smallskip 
    \includegraphics[width=\columnwidth]{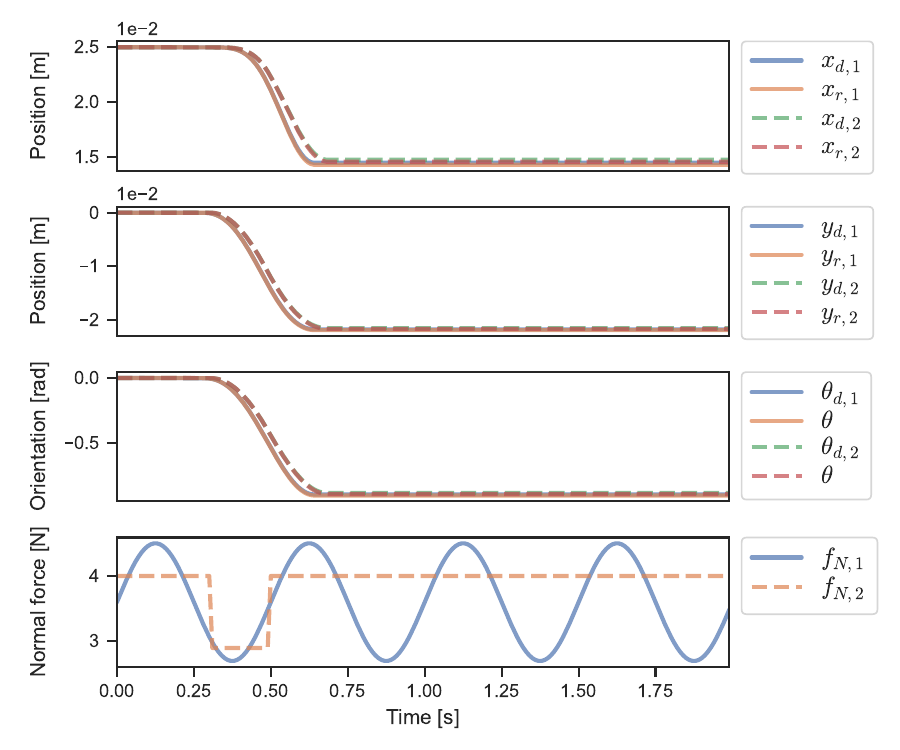}
    \caption{Case 1: circular contact surface ($0.01$ [m] radius) with CoM is placed $25$ $mm$ from CoA on the x-axis. The pose of the CoM is tracked. The subscript number refers to which normal force profile is used ($\mathcolor{blue}{f_{N,1}}$ or $\mathcolor{blue}{f_{N,2}}$).}
    \label{fig:slip_stick_circular_com}
    \vspace*{-0.6cm}
\end{figure}
\begin{figure}
    \centering
    \smallskip 
    \includegraphics[width=\columnwidth]{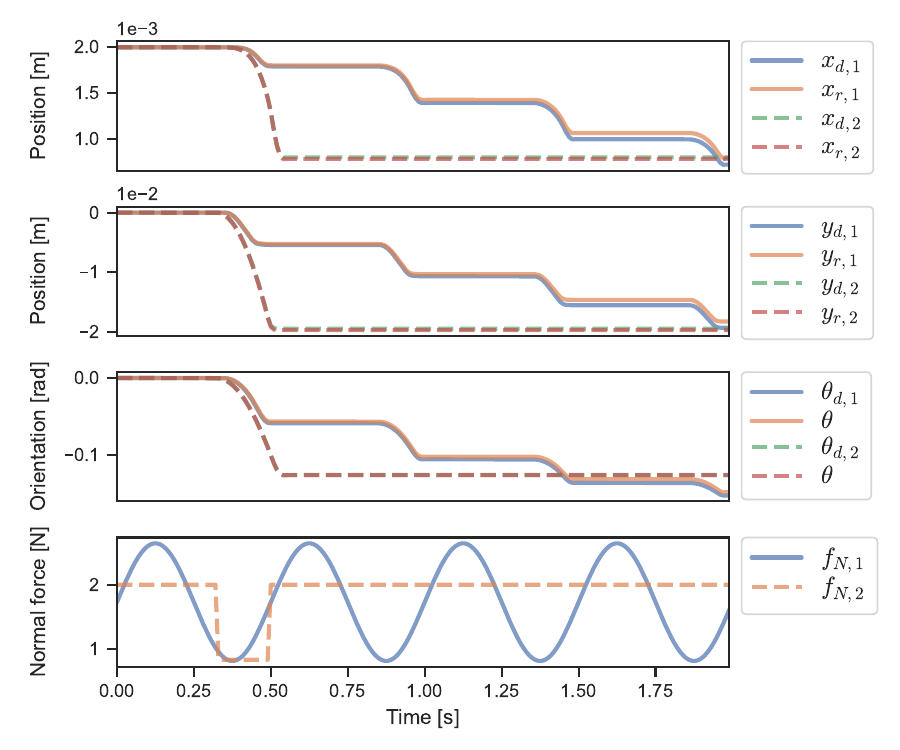}
    \caption{Case 2: simulation with a circular contact surface ($0.01$ [m] radius) when CoM is placed $2$ $mm$ from CoA on the x-axis. The subscript number refers to which normal force profile is used ($\mathcolor{blue}{f_{N,1}}$ or $\mathcolor{blue}{f_{N,2}}$).}
    \label{fig:slip_stick}
    \vspace*{-0.5cm}
\end{figure}

In this section, a simulation is performed on a rectangular object placed in a parallel gripper. The object has the dimensions of $0.15 \times 0.08$ [m] and weighs $0.2$ [kg]. \textcolor{blue}{Two} cases are considered: case 1, where the Center of Mass (CoM) is positioned $25$ [mm] from the Center of Area (CoA) \textcolor{blue}{of the contact surface}, and case 2, where the CoM is placed $2$ [mm] from the CoA, as illustrated in Fig. \ref{fig:slip_stick_view}. For each case, two different normal force profiles are tested, $\mathcolor{blue}{f_{N,1}}$ and $\mathcolor{blue}{f_{N,2}}$ \textcolor{blue3}{(}see Fig. \ref{fig:slip_stick_circular_com} and \ref{fig:slip_stick}\textcolor{blue3}{)}. \textcolor{blue}{The normal force is applied step-wise constant at a frequency of 100 Hz in all experiments. Additionally, two types of normal force dependent contact surfaces are simulated.}  

\begin{figure}
    \centering
    \smallskip 
    \includegraphics[width=\columnwidth]{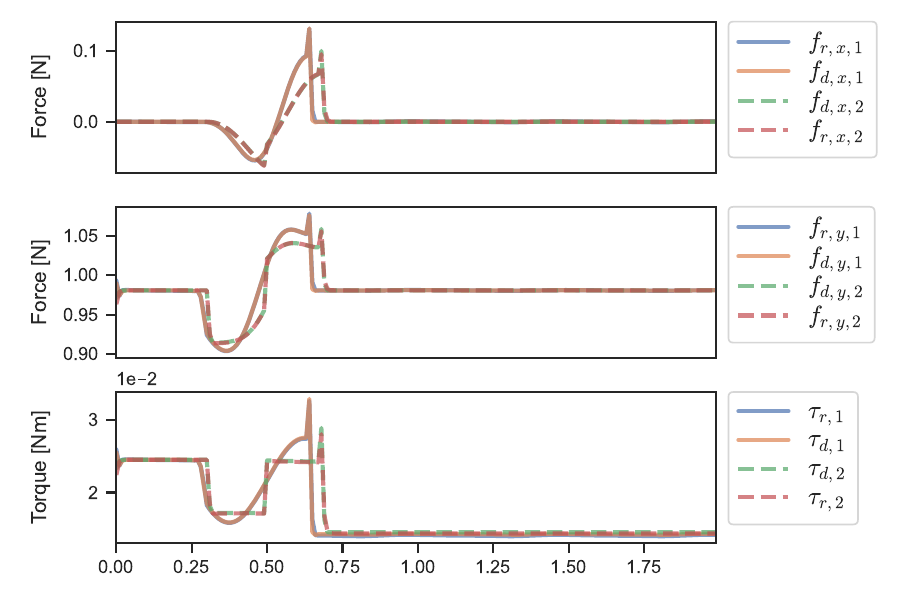}
    \caption{\textcolor{blue}{Case 1: circular contact surface ($0.01$ [m] radius) with CoM is placed $25$ $mm$ from CoA on the x-axis. The subscript number refers to which normal force profile is used ($\mathcolor{blue}{f_{N,1}}$ or $\mathcolor{blue}{f_{N,2}}$).}}
    \label{fig:slip_stick_circular_com_force}
    \vspace*{-0.5cm}
\end{figure}

In case 1, a larger normal force \textcolor{blue}{than for case 2} is required to initiate sticking, and the CoA behaves akin to a pivot point, as depicted in Fig. \ref{fig:slip_stick_a}. Fig. \ref{fig:slip_stick_circular_com} demonstrates that both normal force profiles result in a similar displacement of the CoM. \textcolor{blue}{The forces acting at the \textcolor{blue3}{center} of the contact surface are shown in Fig. \ref{fig:slip_stick_circular_com_force}. The friction force from the reduced model follows the distributed model closely and results in near-identical simulation outcomes.}  


\begin{figure}
    \centering
    \smallskip 
    \includegraphics[width=\columnwidth]{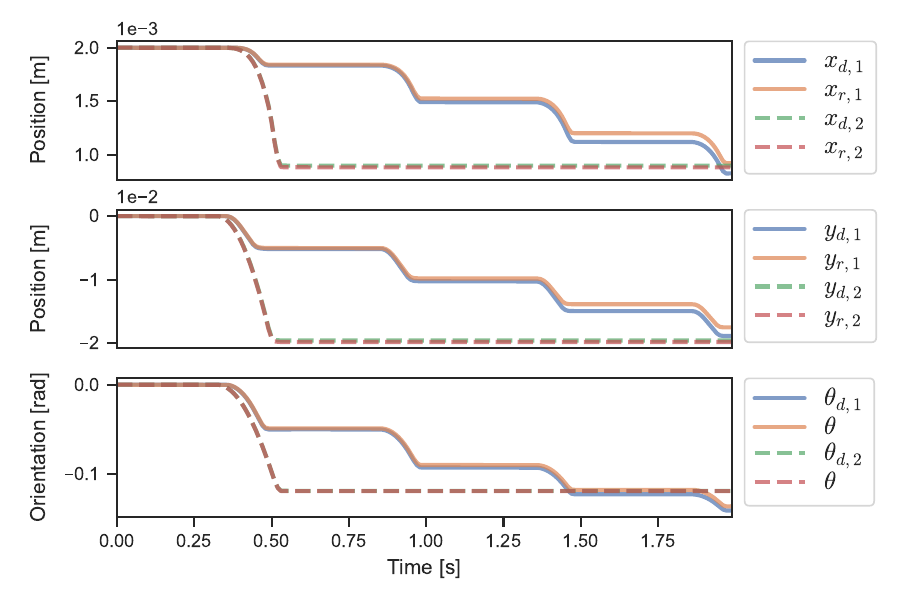}
    \caption{Case 2: simulation from Fig. \ref{fig:slip_stick} but with a $0.02 \times 0.02$ [$\textrm{m}^2$] square contact surface.}
    \label{fig:slip_stick_square}
    \vspace*{-0.5cm}
\end{figure}

\begin{figure}
    \centering
    \smallskip 
    \includegraphics[width=\columnwidth]{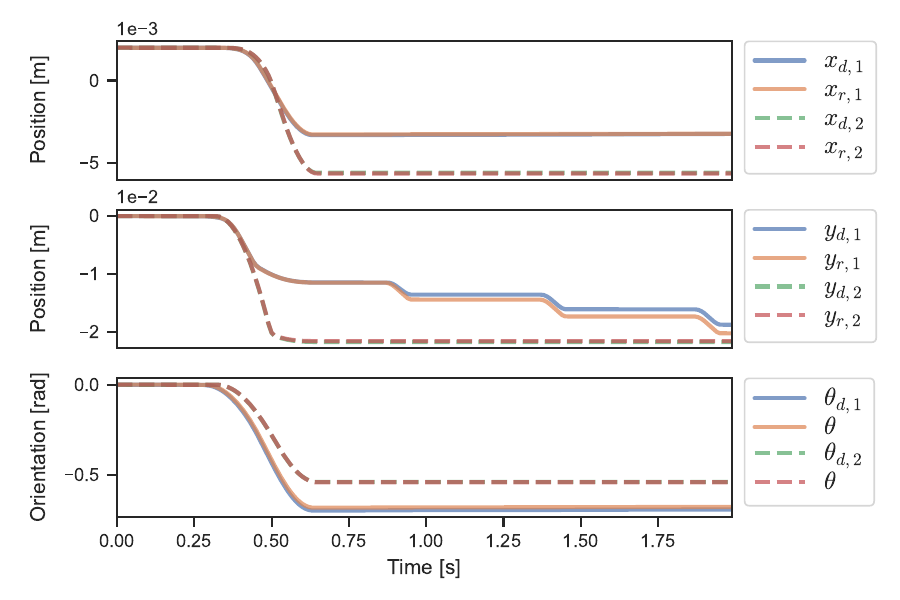}
    \caption{Case 2: simulation from Fig. \ref{fig:slip_stick} but with a $0.02$ [m] gradient line contact surface. The gradient line contact is rotated so the pressure gradient is the largest in $x<0$ and smallest in $x>0$.}
    \label{fig:slip_stick_lineGrad}
    \vspace*{-0.5cm}
\end{figure}

\begin{figure}
    \centering
    \smallskip 
    \includegraphics[width=\columnwidth]{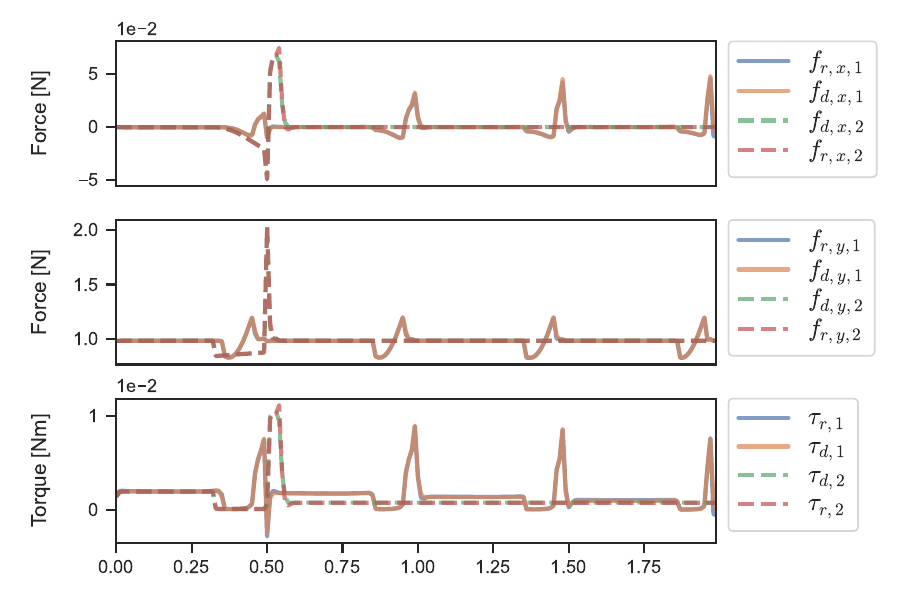}
    \caption{\textcolor{blue}{Case 2: contact forces from Fig. \ref{fig:slip_stick}. The subscript number refers to which normal force profile is used ($\mathcolor{blue}{f_{N,1}}$ or $\mathcolor{blue}{f_{N,2}}$).}}
    \label{fig:slip_stick_force}
    \vspace*{-0.5cm}
\end{figure}

In case 2, there is a more noticeable combined rotational and tangential motion \textcolor{blue3}{(}see Fig. \ref{fig:slip_stick_b}\textcolor{blue3}{)}. The two normal force profiles depicted in Fig. \ref{fig:slip_stick} are tested for three different contact surfaces\textcolor{blue}{: circle, square and gradient line, that are} presented in Fig. \ref{fig:ex_contact_pressure}. The results are shown in Fig. \ref{fig:slip_stick}, \ref{fig:slip_stick_square}, and \ref{fig:slip_stick_lineGrad}. In this case, the choice of the normal force profile influences the object's trajectory \textcolor{blue}{and the resulting forces \textcolor{blue3}{(}see Fig. \ref{fig:slip_stick_force}\textcolor{blue3}{)}}. The circular and square contacts exhibit similar motions, while the gradient line contact results in \textcolor{blue}{a much larger orientation change}. \textcolor{blue}{For the dimensions chosen, the gradient line contact generates less torsional friction than the circle or square contact under similar normal forces.} When $\mathcolor{blue}{f_{N,1}}$ is applied, the object held by the gradient line contact initially slips in both tangential and angular directions, but later only slips in the tangential direction. This occurs because the CoM quickly aligns under the CoP, resulting in no moment generated by gravity.

\begin{figure}
    \centering
    \smallskip 
    \includegraphics[width=\columnwidth]{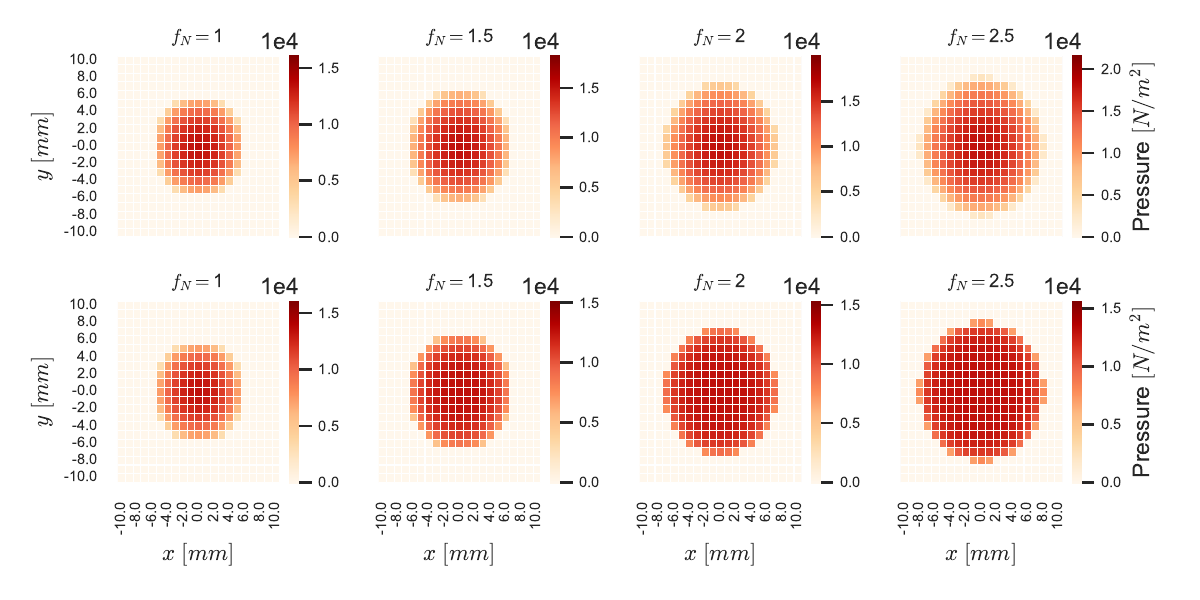}
    \caption{\textcolor{blue3}{Two c}ontact surfaces that are dependent on the normal force. The top surface scales with the normal force and the bottom surface scales ($k=2$) and changes distribution with the normal force ($k=2f_N$).}
    \label{fig:slip_stick_force_prop_surf}
    \vspace*{-0.5cm}
\end{figure}

\begin{figure}
    \centering
    \smallskip 
    \includegraphics[width=\columnwidth]{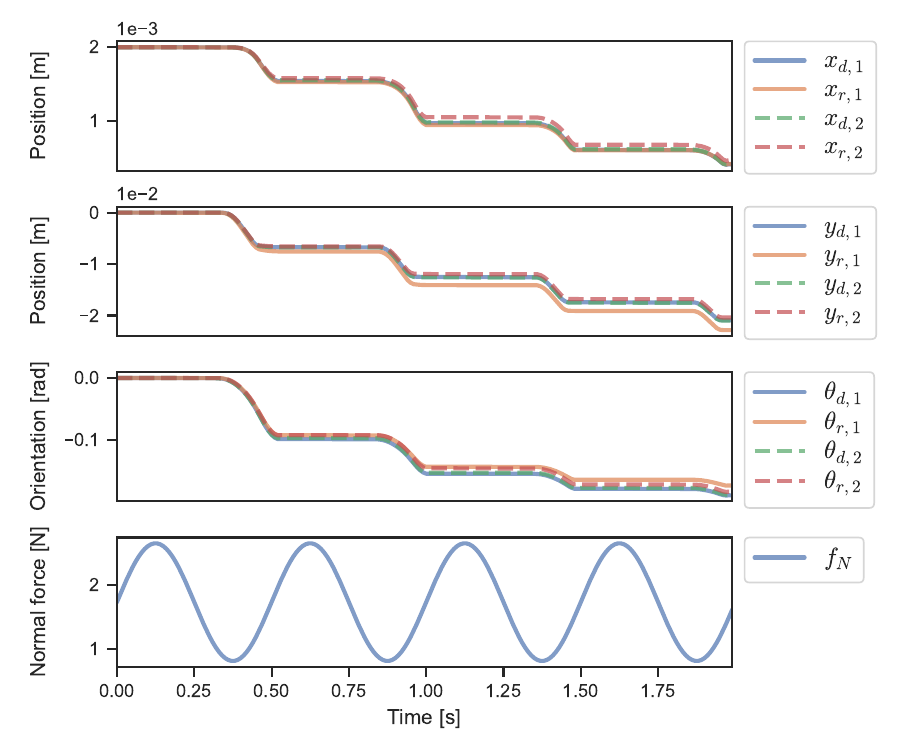}
    \caption{\textcolor{blue}{In-hand slip-stick with normal force-dependent pressure distributions. The subscript number refers to which surface from \ref{fig:slip_stick_force_prop_surf} was used, $1$ is for the top and $2$ for the bottom.}}
    \label{fig:slip_stick_force_prop_pos}
    \vspace*{-0.5cm}
\end{figure}

\begin{figure}
    \centering
    \smallskip 
    \includegraphics[width=\columnwidth]{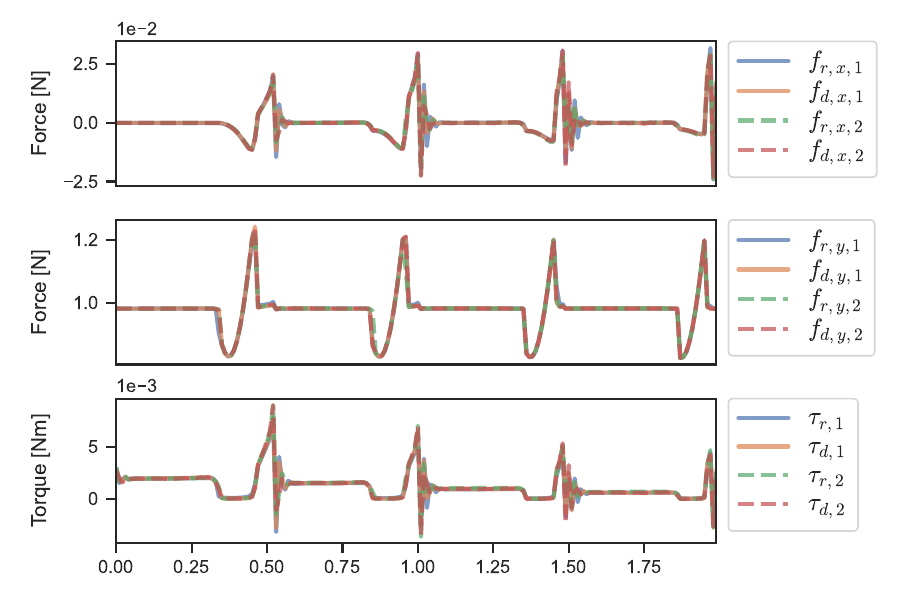}
    \caption{\textcolor{blue}{Friction forces from the in-hand slip-stick experiment in Fig. \ref{fig:slip_stick_force_prop_pos}. The subscript number refers to which surface from Fig. \ref{fig:slip_stick_force_prop_surf} was used, $\cdot_1$ is for the top and $\cdot_2$ for the bottom surface.}}
    \label{fig:slip_stick_force_prop}
    \vspace*{-0.5cm}
\end{figure}

\textcolor{blue}{
Unitl now, the surfaces tested have had a shape that is independent of the normal force. Fig. \ref{fig:slip_stick_force_prop_surf} introduced two contact surfaces that have a normal force-dependent shape and are based on the contact surfaces presented in \cite{xydas1999modeling}. The radius $a$ of the contact surface is modelled as:   
\begin{equation}
    \mathcolor{blue}{a = 6 f_N^{1/3}}
\end{equation}
and the contact pressure for a coordinate is:
\begin{equation}\label{eq:contact_model_p}
\resizebox{0.89\columnwidth}{!}{
    $p(x, y) = \begin{cases}
    1.144 \frac{f_N}{\pi a^2} \left[1 - \left(\frac{||\boldsymbol{\rho}(x, y)||}{a}\right)^k\right]^{1/k} & ||\boldsymbol{\rho}(x, y)|| \leq a \\
    0 &  ||\boldsymbol{\rho}(x, y)|| > a 
    \end{cases}$}
\end{equation}
where $k$ changes the pressure distribution. From Fig. \ref{fig:slip_stick_force_prop_surf}, the top surface has a constant $k=2$ and the surface only scales with the normal force. The bottom surface in the figure has a pressure distribution $k=2f_N$ that changes with the normal force. Fig. \ref{fig:slip_stick_force_prop_pos} shows the experiment with an object in a gripper under varying normal load. Both surfaces give similar outcomes for both friction models. The computational time for this experiment is presented in \textcolor{blue3}{Section} \ref{sec:pre_comp_run_time}. The resulting friction forces can be seen in Fig. \ref{fig:slip_stick_force_prop}.}

\textcolor{blue}{From the slip-stick simulations it can be concluded that the reduced model results in similar simulation outcomes as the distributed model. In the next section, computational times are presented.}

{
\renewcommand{\arraystretch}{1.2}
\begin{table}[pbb]
\scriptsize
\centering
\caption{Simulation time (2 [s]) \label{tab:sim_time}}
\begin{tabular}{c|c|c}
\thickhline
 Model/Surface & Distributed & Reduced \\ \thickhline
 Circle & 217.98 [s]& 3.47 [s]\\ \hline
 $k=2$ & 237.06 [s]& 3.49 [s]\\ \hline
 $k=2f_N$ & 238.95 [s]& 70.23 [s]\\ \hline
\end{tabular}
\end{table}
}

\subsection{Pre-computation and execution time}\label{sec:pre_comp_run_time}
The main advantage of the reduced model is the faster computational time and the reduced number of state variables. The time to simulate 2 seconds for the slip-stick experiments is presented in \textcolor{blue3}{T}able \ref{tab:sim_time}. For the pressure distribution and the object dynamics implemented in Python and the friction model in C++, the reduced model offers approximately 63 times faster solution times than the distributed model. 
A more interesting case is to compare the simulation times for the normal force-dependent pressure distributions. For the surface with $k=2$ the computational cost is only slightly higher than the circle contact, this follows from that the limit surface does not need to be re-calculated. The only re-calculations necessary are $\mathcolor{blue}{r}$ and $u$ from \eqref{eq:r_a} and \eqref{eq:psi} respectively, where $\mathcolor{blue}{r}$ then shifts the output of the limit surface function in \eqref{eq:ls_shift} and scales \eqref{eq:z_rate_red}. For the surface with $k=2f_N$, the limit surface needs to be re-calculated each time the normal force changes, this severely impacts the computational cost of the reduced model. In these simulations the normal force is updated at 100 Hz, increasing the computational time by approximately 20 times. However, the reduced model still outperforms the distributed model. 
\textcolor{blue2}{For in-hand manipulations with soft fingers with axis-symmetric pressure distributions, \cite{costanzo2019two} showed that the exponent in \eqref{eq:contact_model_p} only has a minor impact on the limit surface and can often be modelled as a constant. This corresponds well with our results in Fig. \ref{fig:slip_stick_force_prop_pos} and a constant $k$ allows the simulations to run faster \textcolor{blue3}{(}see \textcolor{blue3}{T}able \ref{tab:sim_time}\textcolor{blue3}{)}}. With the simulation settings in \textcolor{blue3}{T}able \ref{tab:sim_settings} the reduced model has 3 state variables while the distributed has 882 state variables, which the solver has to store and update.

\begin{figure}
    \centering
    \smallskip 
    \includegraphics[width=\columnwidth]{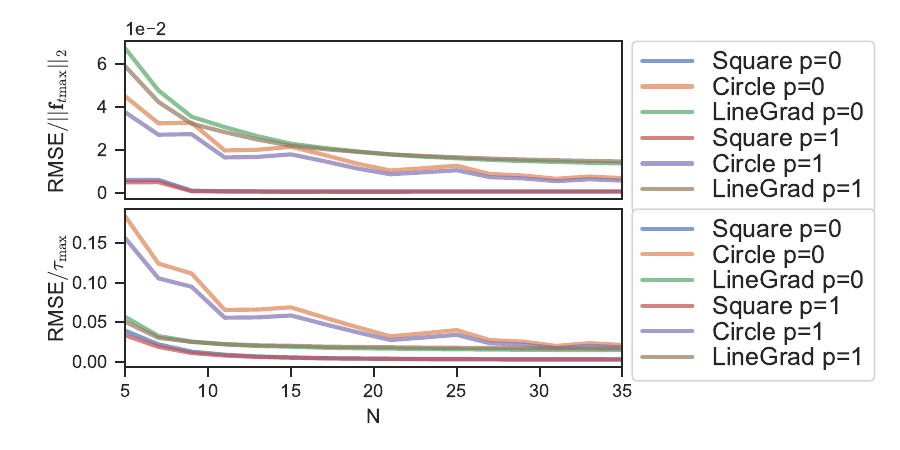}
    \caption{The normalized RMSE for the reduced model over the number of cells ($N \times N$) used during pre-computation. The top plot is for tangential friction and the bottom is for torque. The ground truth is calculated with $101 \times 101$ cells.}
    \label{fig:num_cells_red}
    \vspace*{-0.5cm}
\end{figure}

\begin{figure}
    \centering
    \smallskip 
    \includegraphics[width=\columnwidth]{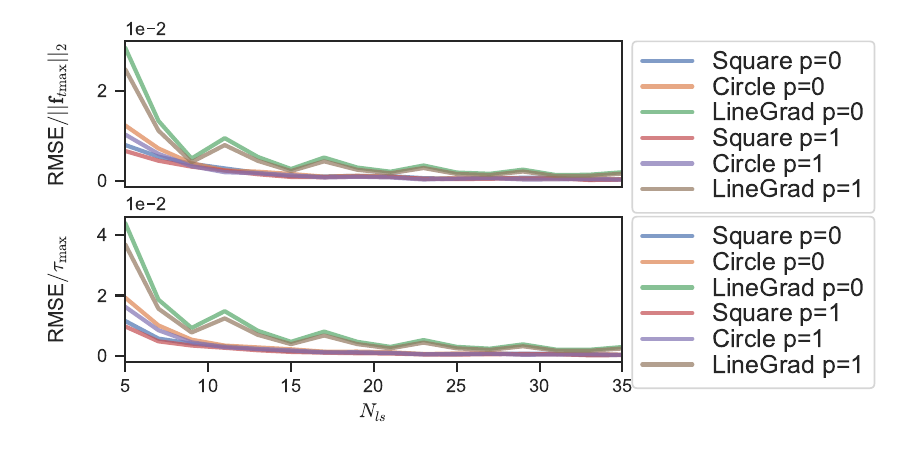}
    \caption{The normalized RMSE for the reduced model over different resolutions $\mathcolor{blue}{N_{\textrm{ls}}}$ of the pre-computed limit surface. The ground truth is calculated with $\mathcolor{blue}{N_{\textrm{ls}}}= 100$, otherwise the settings in \textcolor{blue3}{T}able \ref{tab:sim_settings} are used. }
    \label{fig:num_segments_ls}
    \vspace*{-0.5cm}
\end{figure}

\textcolor{blue}{An experiment akin to the one depicted in Fig. \ref{fig:num_cells} was conducted to explore how the number of cells used during pre-calculation of the limit surface affects the accuracy of the reduced model, \textcolor{blue3}{since} the number of cells is correlated to the mesh-size of the discretization. 
Here the number of cells corresponds to the distributed model used to calculate the limit-surface during pre-computation. The baseline is the planar distributed model computed with $101 \times 101$ cells and the experiment is simulated with a fixed time step of ($\Delta t = 1\textrm{e}{-5}$ \textcolor{blue3}{[s]}).} 
\textcolor{blue3}{The RMSE values from Fig. \ref{fig:num_cells} and \ref{fig:num_cells_red} show that both models are comparable in accuracy.}
The reduced model exhibits a slight advantage due to the bilinear interpolation\textcolor{blue}{, increasing the accuracy slightly by reducing the stepping effect in the forces.}

{
\renewcommand{\arraystretch}{1.2}
\begin{table}[ptt]
\scriptsize
\centering
\caption{ Computational cost (reduced model) \label{tab:comp_time_reduced}}
\resizebox{\columnwidth}{!}{%
\begin{tabular}{c|c|c|c|c|c|c|c|c}
\thickhline
$N$ & 5 & 9  & 13 & 17 & 21 & 25 & 29 & 33\\ \thickhline
 Pre-compute $[s]$ & 2.4e-2 & 5.4e-2 & 1.0e-1 & 1.8e-1 & 2.6e-1 & 3.7e-1 & 4.9e-1 & 6.3e-1  \\ \hline
 Running $\textrm{it}/\textrm{s}$ & 8.7e5 & 8.5e5 & 7.7e5 & 8.6e5 & 8.7e5 &  8.6e5 & 8.6e5 & 8.6e5 \\ \hline
\end{tabular}
}
\end{table}
}
{
\renewcommand{\arraystretch}{1.2}
\begin{table}[ptt]
\scriptsize
\centering
\caption{ Computational cost $\mathcolor{blue}{N_{\textrm{ls}}}$ (reduced model) \label{tab:comp_time_reduced_ls}}
\resizebox{\columnwidth}{!}{%
\begin{tabular}{c|c|c|c|c|c|c|c|c}
\thickhline
$\mathcolor{blue}{N_{\textrm{ls}}}$ & 5 & 9  & 13 & 17 & 21 & 25 & 29 & 33\\ \thickhline
 Pre-compute $[s]$ & 2.2e-2 & 5.8e-2 & 1.1e-1 & 1.9e-1 & 2.9e-1 & 3.9e-1 & 5.3e-1 & 7.0e-1  \\ \hline
\end{tabular}
}
\end{table}
}

The primary motivation behind the reduced model is its computational efficiency, as it offers notable speed advantages irrespective of the number of cells employed, as shown in \textcolor{blue3}{T}able \ref{tab:comp_time_reduced}. For instance, with $21 \times 21$ cells, the reduced model \textcolor{blue}{is} approximately 80 times faster than the distributed model \textcolor{blue3}{(}compare \textcolor{blue3}{T}able \ref{tab:comp_time} and \ref{tab:comp_time_reduced}\textcolor{blue3}{)}. Importantly, this improved speed does not compromise the accuracy significantly, as demonstrated by the similarity between Fig. \ref{fig:num_cells} and \ref{fig:num_cells_red}. The values presented in \textcolor{blue3}{T}ables \ref{tab:comp_time} and \ref{tab:comp_time_reduced} represent averages from 10 runs for each configuration. The pre-computation time demonstrates a linear relationship with the number of cells ($N \times N$), as outlined in \textcolor{blue3}{T}able \ref{tab:comp_time_reduced}. It is worth noting that if the contact shape changes\textcolor{blue}{, i.e. Fig. \ref{fig:slip_stick_force_prop_surf}, some or all} pre-computed variables necessitate recalculation. Additionally, the second factor affecting pre-computation is $\mathcolor{blue}{N_{\textrm{ls}}}$, which characterizes the resolution of the pre-computed limit surface \textcolor{blue3}{(}see \textcolor{blue3}{T}able \ref{tab:comp_time_reduced_ls}\textcolor{blue3}{)}. The total number of sampled cells amounts to $4\mathcolor{blue}{N_{\textrm{ls}}}^2$. Fig. \ref{fig:num_segments_ls} illustrates the normalized RMSE dependency on $\mathcolor{blue}{N_{\textrm{ls}}}$ for the velocity profile depicted in Fig. \ref{fig:circle_dist_lugre}, with the baseline of $\mathcolor{blue}{N_{\textrm{ls}}} = 100$. The observed trend in Fig. \ref{fig:num_segments_ls} indicates rapid convergence to low error values for square and circular contacts, whereas the gradient line contact exhibits slower convergence. This slower convergence can be attributed to the less uniform force distribution when sampled with spherical coordinates, as illustrated in Fig. \ref{fig:limit_surface}.

\section{Conclusion}
In this paper, we have presented a distributed planar friction model that extends the LuGre model and adapts the Elasto-Plastic model \textcolor{blue2}{in order to capture the critical stick-slip regime} for the planar case.
The planar distributed model requires numerical approximations to be solved, which can be computationally intensive. To address this, we have introduced a reduced planar friction model that offers significant computational speedup while maintaining similar dynamics to the distributed model. 
\textcolor{blue2}{For tasks where accurately modelling the slip-stick regime is of critical importance, like controlled in-hand sliding and manipulation of objects, our method proposes an alternative to hybrid friction models and extends what has been developed for one DoF to the planar case.}

Nevertheless, there are still numerous challenges to tackle in friction modelling and the integration of friction models with solvers and object dynamics. In our current approach, the limit surface needs to be re-computed whenever the shape of the contact surface changes. Future research could explore the possibility of generalizing the pre-computation for all contact shapes and devising efficient methods for categorizing and identifying contact shapes. The proposed friction models hold the potential for developing tactile-based perception, in-hand control of sliding objects or object manipulation based on controllable contact surfaces. However, to further advance planar friction models, it would be advantageous to establish an accessible and comprehensive benchmark based on real-world experiments.


\bibliographystyle{ieeetr}
\bibliography{references.bib}

\begin{IEEEbiography}[{\includegraphics[width=1in,height=1.25in,clip,keepaspectratio]{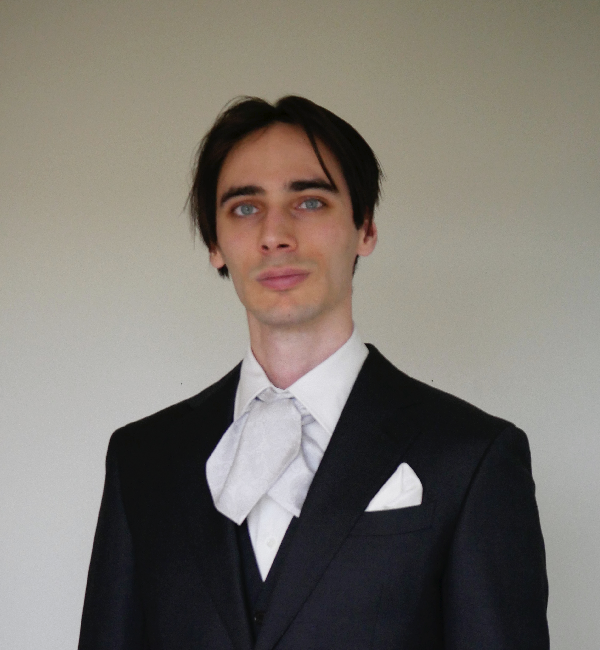}}]{Gabriel Arslan Waltersson}
received the B.S. in Mechanical Engineering and the M.S. degree in Automation and Mechatronics Engineering, in 2019 and 2021, respectively, from Chalmers University of Technology, Gothenburg, Sweden, where he is currently working toward the PhD. degree in Electrical Engineering and is affiliated with WASP.

His research interests include haptic perception for in-hand object tracking, deformable objects and hardware and sensor design. 
\end{IEEEbiography}
\vskip -2\baselineskip plus -1fil
\begin{IEEEbiography}    [{\includegraphics[width=1in,height=1.25in,clip,keepaspectratio]{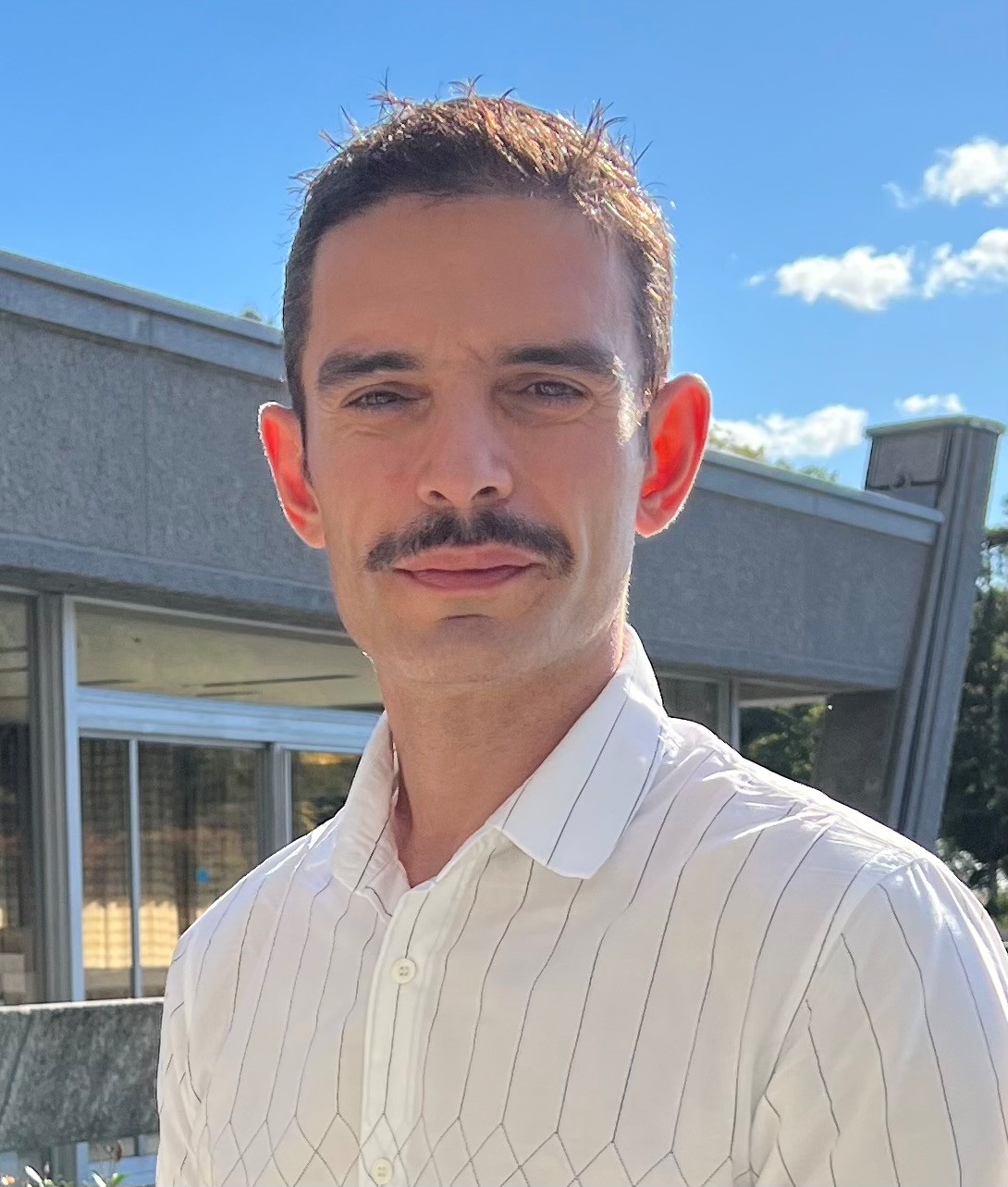}}] {Yiannis Karayiannidis} received a Diploma in Electrical and Computer Eng. (2004) and a Ph.D. degree in Electrical Eng. from Aristotle University of Thessaloniki, Greece (2009).
He is currently an Associate Professor with the Dept. of Automatic Control, Faculty of Engineering at Lund University, Sweden. He is ELLIIT-recruited and  WASP-affiliated faculty. 
He was previously affiliated with KTH, Royal Institute of Technology, Stockholm, Sweden (2011-2020) and Chalmers University of Technology, Gothenburg, Sweden (2015-2022). His research interests include robot control, manipulation and navigation in human-centered environments, dual-arm manipulation, mobile manipulation, force control, haptic perception, robotic assembly, cooperative multi-agent robotic systems, physical human–robot interaction but also adaptive and nonlinear control systems.
\end{IEEEbiography}
\end{document}